\documentclass[11pt]{article}              

\usepackage[margin=25mm]{geometry}
\usepackage{graphicx}
\usepackage{colortbl}
\usepackage{fancyhdr}
\usepackage{amssymb,amsfonts,amsmath,amsthm}
\usepackage{natbib}
\usepackage{bstnotations}
\usepackage{setspace}

\usepackage{authblk}
	
\graphicspath{{./},{./Figures/}}
\usepackage{subfig}
\usepackage[dvipsnames]{xcolor}
\usepackage[flushleft]{threeparttable}
\usepackage{algpseudocode}
\usepackage{algorithm}
\usepackage{algorithmicx}
\newcommand*\Let[2]{\State #1 $\gets$ #2}
\algrenewcommand\algorithmicrequire{\textbf{Initialization:}}
\usepackage[flushleft]{threeparttable}
\usepackage{natbib}
\usepackage{tikz}
\usetikzlibrary{shapes.geometric, arrows,positioning}

\tikzset{
	>=stealth',
	punkt/.style={
		rectangle,
		rounded corners,
		draw=black, very thick,
		text width=6.5em,
		minimum height=2em,
		text centered},
	pil/.style={
		->,
		thick,
		shorten <=2pt,
		shorten >=2pt,}
}

\tikzstyle{optimization} = [rectangle, rounded corners, minimum width=0.5cm, minimum height=1cm, text centered,text width=6cm, draw=black]

\tikzstyle{counter} = [ellipse, minimum width=1.5cm, minimum height=1cm, text centered, draw=black]

\tikzstyle{decision} = [diamond, minimum width=0.5cm, minimum height=0.5cm, text centered, draw=black]

\tikzstyle{Fin} = [rectangle, rounded corners, minimum width=1cm, minimum height=1cm, text centered,text width=1cm, draw=black]

\tikzstyle{line} = [draw, -latex']
\tikzstyle{arrow} = [thick,->,>=stealth]

\tikzstyle{stop} = [rectangle, rounded corners, minimum width=1cm, minimum height=1cm,text centered, draw=black]

\tikzstyle{init} = [rectangle, rounded corners, minimum width=0.5cm, minimum height=1cm, text centered,text width=8cm, draw=black]

\tikzstyle{glo_enr} = [rectangle, rounded corners, minimum width=0.5cm, minimum height=1cm,text centered,text width=8cm, draw=black]

\newcommand{\equaref}[1]{Eq.~(\ref{#1})}

\newcommand{\fz}{f_{\boldsymbol{Z}}}
\newcommand{\ft}{f_{\boldsymbol{W}}}
\newcommand{\bd}{{\mathbb D}}

\newcommand{\br}{{\mathbb R}}

\newcommand{\fc}{{\mathfrak c}}

\graphicspath{{./},{./Figures/}}

\begin{document}
\title{Quantile-based optimization under uncertainties using adaptive Kriging surrogate models} 

\author[1,2,3]{M. Moustapha} \author[1]{B. Sudret} \author[2]{J.-M. Bourinet}  \author[3]{B. Guillaume} 

\affil[1]{Chair of Risk, Safety and Uncertainty Quantification,
  
  ETH Zurich, Stefano-Franscini-Platz 5, 8093 Zurich, Switzerland}

\affil[2]{Institut Pascal, Sigma Clermont, CNRS UMR 6602, Aubi\`ere, France}

\affil[3]{PSA Group, Centre Technique de V\'elizy, V\'elizy-Villacoublay, France}
\date{}
\maketitle

\abstract{Uncertainties are inherent to real-world systems. Taking them into
	account is crucial in industrial design problems and this might be
	achieved through reliability-based design optimization (RBDO)
	techniques. In this paper, we propose a quantile-based approach to
	solve RBDO problems.  We first transform the safety constraints
	usually formulated as admissible probabilities of failure into
	constraints on quantiles of the performance criteria. In this formulation, the quantile level controls the degree of conservatism of the design. Starting with the
	premise that industrial applications often involve high-fidelity and
	time-consuming computational models, the proposed approach makes use
	of Kriging surrogate models (a.k.a.  Gaussian process modeling).
	Thanks to the Kriging variance (a measure of the local accuracy of the
	surrogate), we derive a procedure with two stages of enrichment of the
	design of computer experiments (DoE) used to construct the surrogate
	model.  The first stage globally reduces the Kriging epistemic
	uncertainty and adds points in the vicinity of the limit-state
	surfaces describing the system performance to be attained.  The second
	stage locally checks, and if necessary, improves the accuracy of the
	quantiles estimated along the optimization iterations.  Applications
	to three analytical examples and to the optimal design of a car body
	subsystem (minimal mass under mechanical safety constraints) show the
	accuracy and the remarkable efficiency brought by the proposed
	procedure.  \\[1em] 

}

\maketitle

%%%%%%%%%%%%%%%%%%%%%%%%%%%%%%%%%%%%%%%%%%%%%%%%%%%%%%%%%%%%%%%%%%%%%%%%%%%%
\section{Introduction}\label{intro}
In engineering design, one often seeks to lower the product cost while
ensuring its integrity. These are by construction two conflicting
objectives. Optimization has therefore been used as an automatic
procedure to find a good trade-off. The optimal solution
usually lies at the boundary of the feasible space. However
uncertainties are ubiquitous to engineering systems whether arising from
modeling approximations or input parameters inherent variability. They
make any optimal design  solution likely to depart from its real-world
counterpart. Such discrepancy may turn a feasible solution into an
unfeasible one. It is therefore of prime importance to account for
uncertainties during optimization. This is generally achieved through
\emph{robust} and \emph{reliability-based design optimization}
(respectively RDO and RBDO). In the former, emphasis is put on the cost
function. The designer actually searches for a design that is immune to
the inputs uncertainties. The cost function is in this case replaced by
robustness measures which include worst-case scenarios or moment-based
criteria \citep{Trosset1997}. \citet{Beyer2007} and
\citet{Baudoui2012} give a comprehensive review of such techniques. On
the other hand, reliability-based design optimization rather seeks to
balance the cost and the safety requirements by moving the solution away
from the boundary of the admissible space. The work presented in this
paper is concerned with the latter approach.

Following the notations in \citet{Dubourg2011}, a reliability-based design optimization may be formulated as follows:
\begin{equation}\label{eq:001}
\begin{split}
& \ve{d}^\ast = \arg \min_{\ve{d} \in \bd} \fc \prt{\ve{d}} \quad \text{subject to: }
\left\{ \begin{array}{ll}
\mathfrak{f}_j \prt{\ve{d}} \leq 0, \quad &  \acc{j = 1, \ldots, n_s}, \\
\Prob{\mathfrak{g}_k \prt{\ve{X}\prt{\ve{d}},\ve{Z}} \leq 0} \leq \bar{P}_{f_k}, \quad & \acc{k = 1, \ldots, n_h},
\end{array} \right.
\end{split}
\end{equation}
where a cost function $\fc$ is minimized with respect to design
variables $\ve{d}$. This minimization task is carried out under a set of
constraints divided into two groups respectively denoted by \emph{soft}
and \emph{hard} constraints. The $n_s$ soft constraints $\mathfrak{f}_j$
are simple analytical functions, often bounding the design space while
the $n_h$ hard constraints $\mathfrak{g}_k$ are actually the system
performance functions. They rely on the mechanical model $\mathcal{M}_k$
used to predict the structural behavior. In our case, they result from a
finite element model and may be written as $\mathfrak{g}_k = \bar{\mathfrak{g}}_k
- \mathcal{M}_k$, where $\bar{\mathfrak{g}}_k$ is a threshold not to be
exceeded by the structural response which is computed from a simulation model $\ve{x} \mapsto \mathcal{M}_k\prt{\ve{x}}$ (usually a time-consuming finite element model). When safety requirements are of
interest, performance may be measured in terms of a failure probability.
To this end, random variables accounting for the uncertainties in the
inputs are introduced and denoted respectively by $\ve{X} \sim \fx$ for
the design variables and $\ve{Z} \sim \fz$ for the environmental
variables. The former notation means that the distribution of $\ve{X}$
is conditioned on the design parameters. Typically, design parameters
$\ve{d}$ are nominal dimension and $\fx$ models the uncertainties due to
manufacturing tolerances. Environmental variables $\ve{Z}$ may for
instance be parameters of the crash protocol such as the impact speed in
crashworthiness design. By propagating these uncertainties to the
output, the failure probability for a given design $\ve{d}$ reads:
\begin{equation}\label{eq:002}
P_{f_k}\prt{\ve{d}} = \Prob{\mathfrak{g}_k\prt{\ve{W}} \leq 0} = \int_{\mathfrak{g}_k\prt{\ve{w}} \leq 0} \ft\prt{\ve{w}} d\ve{w},
\end{equation}
where $\ve{W} = \acc{\ve{X}|\ve{d},\ve{Z}}^T \sim \ft$ is a vector
gathering all the input parameters of the mechanical model that governs
the structure's behavior. The solution of this integral is generally not
tractable because the failure domain defined by $\acc{\ve{w}:
	\mathfrak{g}_k \prt{\ve{w}} \leq 0}$ has an implicit definition. One
rather resorts to \emph{approximation} or \emph{simulation} methods
\citep{Madsen1986}. Within the former group, the first-order reliability
method (FORM) is the most widely used
\citep{Ditlevsen1996,Lemaire2007,Hasofer1974}. It consists in mapping the
random variables into the standard normal space where the limit-state
surface is linearly approximated. Therefore, the failure probability can
be equivalently expressed by the so-called \emph{reliability index}
\citep{Hasofer1974}. Curvatures of the limit-state surface may be handled
by the second-order reliability method (SORM). As for the simulation
methods, the most straightforward one is crude Monte Carlo simulation
(MCS) where the failure probability is estimated by the relative
occurrence of failed samples. The accuracy of the estimate depends on
the number of samples. For extremely small probabilities of failure, the
required number of samples for an accurate estimate becomes relatively
high, typically $10^{6-8}$, which makes the approach not affordable.
Variance-reduction techniques have been introduced in order to by-pass
this limitation \citep{Asmussen2007}. Applied to rare events simulations,
such techniques include importance sampling \citep{Au1999,Melchers1989}
and subset simulation \citep{Au2001}. The former proceeds by sampling
from an instrumental distribution which puts a higher weight to the
failure domain and afterwards correct the introduced bias appropriately.
The latter splits the failure domain into nested auxiliary domains such that the
failure probability can be estimated by the product of larger ones, the
latter being easier to evaluate by simulation. However, the
computational cost of all these techniques is in the order of $10^{3-4}$
(for each design $\ve{d}$), and can thus not be used within an
optimization loop.

Indeed, the solution of the RBDO problem relies on the estimate of the
failure probability for different values of the design parameters. Many
techniques exist and may be classified into \emph{two-level},
\emph{mono-level} and \emph{decoupled} approaches
\citep{Chateauneuf2008,Aoues2010}. Two-level approaches, which basically
consist of two nested loops, are among the most straightforward to
implement. The outer loop explores the design space and the inner one
solves the reliability analysis for any given design. Usually, the inner
loop resorts to FORM approximations as in the so-called \emph{reliability index} \citep{Enevoldsen1994} and \emph{performance measure} \citep{Tu1997,Tu1999} approaches (respectively RIA and PMA). Simulation techniques may also be used in the inner loop, as we show in the sequel. The mono-level approach transforms the double-loop problem into a single-loop one by introducing optimality criteria for the FORM problem. \citet{Kuschel1997} propose an equivalent formulation based on RIA while \citet{Agarwal2007} rather rely on PMA. Finally, the decoupled approaches transform the double-loop into a sequence of deterministic problems. A well-known example is \emph{sequential optimization and reliability analysis} (SORA) proposed in \citet{Du2004}.  

As introduced above, all theses methods rely on repeated evaluations of
the mechanical models, \ie during the outer optimization loop and more
intensively in the reliability analysis steps. This limits their range
of applications to engineering problems of practical interest. This
issue is even more dramatic in the design of complex industrial systems
which relies on high-fidelity models and henceforth time-consuming
simulations. \emph{Surrogate modeling}, a technique in which the
mechanical model is replaced by a well calibrated easy-to-evaluate
analytical function, has been extensively used in the past decade to
alleviate the computational burden. For instance, support vector
machines have been used for structural reliability assessment in
\citet{Hurtado2001,Bourinet2011,Deheeger2007}. Polynomial chaos expansion
were considered in \citet{BlatmanThesis,Blatman2010,Hu2011}.
Kriging (a.k.a. Gaussian process modeling) has been successfully used for reliability analysis in
\citet{Echard2011,Picheny2010,Bichon2008,Balesdent2013}. For the specific task of RBDO, conservative surrogate models which rely on Kriging or polynomial response surfaces were considered in \citet{Viana2010,Picheny2008}. From another perspective, \citet{Dubourg2011,Chen2015,Lee2011,Li2016}, for instance, have proposed some approaches which rely on locally or globally refined Kriging approximations. Likewise, the present work considers Kriging because it provides 
not only an approximation of the original mechanical model but also
gives a built-in error estimate. This enables adaptive techniques that
further reduce the computational cost.

In contrast to most of the literature in RBDO, we are not interested in
this paper in highly reliable designs, for which the probability of
failure has to be computed by one of the methods mentioned above. Our
goal is rather to develop a \emph{conservative} optimal design
methodology. To this aim, we first introduce a quantile-based design
optimization procedure, in which the hard constraints are formulated on
quantiles of the performance criteria, instead of target probabilities
of failure. At each iteration of the design optimization, we therefore
evaluate quantiles through Monte Carlo simulation instead of solving a
reliability problem. This approach is justified by the "degree of
conservatism" targeted in our applications to car body mass
optimization: $95 \%$-quantiles are indeed considered as sufficiently
conservative, which makes their evaluation relatively easy. As the
performance criteria are obtained from time-consuming simulation (\eg
frontal impact of a full car body or a subsystem), quantile evaluation
must rely on surrogate models. In this respect, the goal of the paper is
to propose a quantile-based design optimization methodology which is
relying on adaptive Kriging surrogate models.

The paper is organized as follows: in Section 2 we introduce the
quantile-based optimization and prove its formal equivalence with the
classical RBDO setting. The basics of Kriging are then summarized in
Section 3. In Section 4, an original two-stage strategy of enrichment of
the experimental designs used in Kriging is proposed, as a means to
reduce the overall computational burden (\eg, at most a few hundreds
runs of the time-consuming computational model) to regions of the design
space that are relevant for optimization. Finally, Section 5 presents
four examples: the three first involve analytical constraints and allows
us to validate our approach against benchmark results. The final example
is a real case study which deals with the mass optimization of a car body subsystem under
crashworthiness constraints.

\section{Formulation of the quantile-based optimization procedure}
\subsection{Equivalence between RBDO and quantile-based
	formulation} Prior to formulating the quantile-based procedure, let
us consider the reliability-based design optimization problem in
\equaref{eq:001}.  By explicitly introducing the computational model
of interest $\mathcal{M}$ which describes the system performance, the
following equivalence holds: \begin{equation}\label{eq:003}
\begin{split}
\Prob{\mathfrak{g} \prt{\ve{X}\prt{\ve{d}},\ve{Z}} \leq 0} \leq \bar{P}_{f} & \Leftrightarrow \Prob{\mathcal{M} \prt{\ve{X}\prt{\ve{d}},\ve{Z}} \geq \bar{\mathfrak{g}}} \leq \bar{P}_{f}, \\
& \Leftrightarrow \Prob{\mathcal{M}
	\prt{\ve{X}\prt{\ve{d}},\ve{Z}} \leq \bar{\mathfrak{g}}} \geq 1
- \bar{P}_{f}, \end{split} \end{equation} where
$\bar{\mathfrak{g}}$ is an upper threshold on the system mechanical
response.

From the last expression, we can introduce the following quantile as an
alternative way of measuring the failure probability:
\begin{equation}
\label{eq:004}
Q_{\alpha} \prt{\ve{d}; \mathcal{M}\prt{\ve{X}\prt{\ve{d}},\ve{Z}}} = \inf \acc{q \in \br\;  : \; \Prob{\mathcal{M} \prt{\ve{X}\prt{\ve{d}},\ve{Z}} \leq q } \geq \alpha},
\end{equation}
where $\alpha = 1 - \bar{P}_{f}$.

The computed quantile may henceforth be used as a measure of reliability given a target failure probability. Considering \equaref{eq:003} and  \equaref{eq:004}, the following equivalence holds:
\begin{equation}
\label{eq:3001}
\Prob{\mathfrak{g} \prt{\ve{X}\prt{\ve{d}},\ve{Z}} \leq 0} \leq
\bar{P}_{f} \Leftrightarrow Q_{\alpha} \prt{\ve{d};
	\mathcal{M}\prt{\ve{X}\prt{\ve{d}},\ve{Z}}}  \leq
\bar{\mathfrak{g}}, 
\end{equation}
where the value of $\alpha$ is directly related to the target failure
probability.

\begin{figure}[!ht]
	\begin{center}
          \subfloat[Safe design: $\Prob{\mathcal{M}
            \prt{\ve{X}\prt{\ve{d}},\ve{Z}} \geq \bar{\mathfrak{g}}}
          \leq \bar{P}_{f} \Leftrightarrow Q_{\alpha} \prt{\ve{d};
            \mathcal{M}\prt{\ve{X}\prt{\ve{d}},\ve{Z}}} \leq
          \bar{\mathfrak{g}}$]{\includegraphics[width=0.8\textwidth]{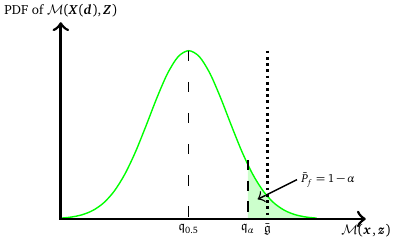}\label{fig:Safe_D}}%
          \\%
          \subfloat[Unsafe design: $\Prob{\mathcal{M}
            \prt{\ve{X}\prt{\ve{d}},\ve{Z}} \geq \bar{\mathfrak{g}}} >
          \bar{P}_{f} \Leftrightarrow Q_{\alpha} \prt{\ve{d};
            \mathcal{M}\prt{\ve{X}\prt{\ve{d}},\ve{Z}}} >
          \bar{\mathfrak{g}}
          $]{\includegraphics[width=0.8\textwidth]{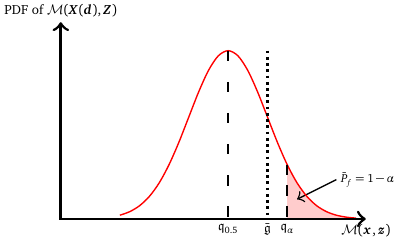}\label{fig:Unsafe_D}}%
		\caption{Comparison of a safe and an unsafe design with respect to a quantile $Q_\alpha$ corresponding to a target failure probability $\bar{P}_f = 1 - \alpha$.} %
		\label{fig:001}%
	\end{center}
\end{figure}

This equivalence between the failure probability and the quantile
estimation is illustrated in \figref{fig:001} where the distributions of
a mechanical response in two configurations are shown. In the upper
panel, the quantile corresponding to the target failure probability is
lower than the constraint threshold. This corresponds to a safe design
since the probability that $\mathcal{M}\prt{\ve{X}\prt{\ve{d}}, \ve{Z}}$
is greater that $\bar{\mathfrak{g}}$ is smaller than $\bar{P}_f = 1 -
\alpha$. In contrast, the lower panel shows an unsafe design since the
associated quantile is higher than the threshold $\bar{\mathfrak{g}}$,
meaning that the probability of failure $\Prob{\mathcal{M}
	\prt{\ve{X}\prt{\ve{d}},\ve{Z}} \geq \bar{\mathfrak{g}} }$ is greater
than $\bar{P}_f$.

Following the previous developments, the RBDO problem of \equaref{eq:001} may eventually be recast as:
\begin{equation}\label{eq:005}
\begin{split}
& \ve{d}^\ast = \arg \min_{\ve{d} \in \bd} \fc \prt{\ve{d}} \quad \text{subject to: }
\left\{ \begin{array}{ll}
\mathfrak{f}_j \prt{\ve{d}} \leq 0, \quad &  \acc{j = 1, \ldots, n_s}, \\
Q_{\alpha_k}\prt{\ve{d}; \mathcal{M}_k\prt{\ve{X}\prt{\ve{d}},\ve{Z}}} \leq \bar{\mathfrak{g}}_k, \quad & \acc{k = 1, \ldots, n_h},
\end{array} \right.
\end{split}
\end{equation}
where $\alpha_k = 1 - \bar{P}_{f_k}$.

\subsection{Monte Carlo estimate of the quantile}
To solve the optimization problem in \equaref{eq:005}, the quantile must be estimated in each iteration for the current design $\ve{d}^{(i)}$. In this paper, we consider crude Monte Carlo sampling. The following steps describe the numerical procedure:
\begin{enumerate}
	\item Sample the Monte Carlo set needed to evaluate the quantile:
	\begin{equation}\label{eq:006}
	\mathfrak{C}_q\prt{\ve{d}^{(i)}} = \acc{\prt{\ve{x}^{(j)},\ve{z}^{(j)}}, j = 1, \ldots, N},
	\end{equation}
	where $\ve{X} \sim f_{\ve{X}|\ve{d}^{(i)}}$, $\ve{Z} \sim \fz$ and $N$ is the size of the Monte Carlo sample set.
	\item Compute the set of associated responses for each mechanical model:
	\begin{equation}
	\ve{\mathcal{Y}}_k = \acc{y_k^{(j)} = \mathcal{M}_k\prt{\ve{x}^{(j)},\ve{z}^{(j)}}, j = 1, \ldots, N}
	\end{equation}
	\item Sort them in ascending order such that $y_{k_{(1)}} \leq y_{k_{(2)}} \leq \ldots \leq y_{k_{(N)}}$
	\item Retrieve the quantile corresponding to the $k$-th constraint by:
	\begin{equation}\label{eq:008}
	Q_{\alpha_k} \prt{\ve{d}^{(i)}; \mathcal{M}_k\prt{\ve{X}\prt{\ve{d}^{(i)}},\ve{Z}}} \equiv \mathfrak{q}_{\alpha_k} \prt{\ve{d}^{(i)}} = y_{k_{(\lfloor N \ve{\alpha}_k \rfloor)}},
	\end{equation}
	where $\lfloor t \rfloor$ denotes the floor function yielding the largest integer smaller than $t$.	
\end{enumerate}

To apply this approach, the Monte Carlo sample set in \equaref{eq:006} needs to be large enough so that the computed quantile is accurate. For our application, where the target failure probability is $1 \% - 10 \%$, we choose $N = 10,000$. As this simulation is embedded in the iterative process of optimization, the number of calls to the mechanical model may reach hundreds of thousands. When a high-fidelity model is involved, such a large number of calls is not affordable. We therefore couple the proposed approach to a well-known surrogate modeling technique, namely Kriging.

\section{Kriging (a.k.a Gaussian process modeling)}
Surrogate models have been increasingly used as proxies of time-consuming functions in the past decade. In the computer experiments setting, such a function is considered to be a black-box \ie only pairs of inputs/outputs are known with respect to a limited set of observations. This set constitutes the \emph{design of experiments} and reads, for a given model $\mathcal{M}$:
\begin{equation}\label{eq:009}
\mathcal{D} = \acc{\prt{\ve{x}_i, y_i}, \ve{x}_i \in \mathbb{R}^s, y_i = \mathcal{M}\prt{\ve{x}_i}, i = 1, \ldots, n},
\end{equation}
where $\ve{x}_i$ is an $s$-dimensional input, $y_i$ is the corresponding scalar output and $n$ is the number of available observations in the design of experiments.

Kriging a.k.a. Gaussian process modeling \citep{Santner2003} is one particular emulator which considers the function $\mathcal{M}$ to approximate as a realization of a stochastic process, which may be cast as:
\begin{equation}
\mathcal{M}\prt{\ve{x}} = \sum_{j=1}^{p} \beta_j f_j\prt{\ve{x}} + Z \prt{\ve{x}},
\end{equation}  
where the first summand is the deterministic part referred to as the \emph{trend}. It reads as a linear combination of a vector of $p$ weight coefficients $\ve{\beta} = \acc{\beta_j, j = 1, \ldots, p}$ and a set of function basis $\ve{f} = \acc{f_j, j = 1, \ldots, p}$. The second summand is a zero-mean stationary Gaussian process. It is completely defined by its auto-covariance function $\Cov{Z\prt{\ve{x}},Z\prt{\ve{x}'}} = \sigma^2 R \prt{\ve{x},\ve{x}'; \ve{\theta}}$, where $\sigma^2$ is the constant variance of the Gaussian process, $R$ is the auto-correlation function whose hyperparameters are gathered in the vector $\ve{\theta}$.

The calibration of the Kriging model involves making a few choices that can be motivated by some prior knowledge on the function to approximate. The first one is the choice of the mean trend. In this work, we consider an unknown constant trend. This results in the so-called \emph{ordinary Kriging}. The second one is the choice of the auto-correlation function which encodes assumptions such as the degree of regularity of the underlying process. A wide family of auto-correlation functions have been used in the literature. Here, we consider the Mat\'ern $5/2$ auto-correlation family, defined in the one-dimensional case by:
\begin{equation}\label{eq:101}
R\prt{x,x';l} = \prt{1 + \sqrt{5} \frac{\abs{x-x'}}{l} + \frac{5}{3} \frac{\prt{x-x'}^2}{l^2}} \exp \prt{-\sqrt{5} \frac{\abs{x-x'}}{l}},
\end{equation}
where $l$ is the so-called \emph{characteristic length scale}. The multi-dimensional case is obtained by tensor product of the above equation:
\begin{equation}\label{eq:102}
R\prt{\ve{x},\ve{x}';\ve{\theta}} = \prod_{i=1}^{s}	R\prt{x_i,x^{'}_{i};\theta_i}.
\end{equation}

Once these choices are made, the Kriging predictor at the point $\ve{x}$ is assumed to follow a normal distribution $\widehat{\mathcal{M}}\prt{\ve{x}} \sim \mathcal{N}\prt{\mu_{\widehat{\mathcal{M}}}\prt{\ve{x}}, \widehat{\sigma}^2_{\widehat{\mathcal{M}}}\prt{\ve{x}}}$:
\begin{equation}\label{eq:389}
\begin{split}
& \mu_{\widehat{\mathcal{M}}}\prt{\ve{x}} = \ve{f}^T \prt{\ve{x}} \widehat{\ve{\beta}} + \ve{r}^T\prt{\ve{x}} \mat{R}^{-1} \prt{\ve{y} - \mat{F}^T \widehat{\ve{\beta}}},\\
& \widehat{\sigma}^2_{\widehat{\mathcal{M}}}\prt{\ve{x}} = \sigma^2 \prt{1 - \ve{r}^T\prt{\ve{x}} \mat{R}^{-1} \ve{r}\prt{\ve{x}} + \ve{u}^T\prt{\ve{x}} \prt{\mat{F}^T \mat{R}^{-1}\mat{F}}^{-1} \ve{u}\prt{\ve{x}}},
\end{split}
\end{equation}
where $\widehat{\ve{\beta}} = \prt{\mat{F}^T \mat{R}^{-1} \mat{F}}^{-1} \mat{F}^T \mat{R}^{-1} \ve{y}$ is the generalized least-square estimate of the weight coefficients $\ve{\beta}$, $\ve{r}\prt{\ve{x}}$ is a vector of cross-correlations between the point $\ve{x}$ and each point of the design of experiments, $\mat{F}$ is the information matrix whose components are $f_j\prt{\ve{x}_i}, i = \acc{1, \ldots, n}, j = \acc{1, \ldots, p}$ and $\ve{u} = \mat{F}^T \mat{R}^{-1} \ve{r}\prt{\ve{x}} - \ve{f}\prt{\ve{x}}$ has been introduced for the sake of clarity.
Beside the prediction given by $\mu_{\widehat{\mathcal{M}}}\prt{\ve{x}}$, Kriging features a measure its own accuracy through the prediction variance $\widehat{\sigma}^2_{\widehat{\mathcal{M}}}\prt{\ve{x}}$. Confidence intervals on the prediction can then be derived since the distribution of the prediction $\widehat{\mathcal{M}}\prt{\ve{x}}$ is Gaussian by assumption. More importantly, this has supported the development of infill-sampling criteria used for the adaptive refinement of Kriging models. 

Eventually, one has to estimate the hyperparameters of the auto-correlation functions to completely define the Kriging predictor. This is achieved through automatic calibration following techniques such as \emph{cross-validation} or \emph{maximum likelihood estimation}. The latter is used in this work and boils down to the following optimization problem:
\begin{equation}\label{eq:012}
\widehat{\ve{\theta}} = \arg \min_{\ve{\theta} \in \br^d} \psi \prt{\ve{\theta}} = \widehat{\sigma^2}\prt{\ve{\theta}} \det \mat{R}\prt{\ve{\theta}}^{\frac{1}{n}},
\end{equation}
where $\psi \prt{\ve{\theta}}$ is the so-called \emph{reduced likelihood} function and $d$ is the number of parameters in $\ve{\theta}$ \citep{Koehler1996,DubourgThesis}.

The accuracy of the solution of the optimization problem in \equaref{eq:012} is crucial as it conditions the quality of the Kriging predictor. General-purpose algorithms such as genetic algorithm or BFGS are often used. Available softwares such as DiceKriging \citep{Roustant2012} in R, UQLab \citep{MarelliSudret2014,Lataniotis2015} or ooDace \citep{Couckyut2013} in MATLAB make use of such algorithms  and more generally provide a framework to train a Kriging model.

\section{Kriging-based optimization}

\subsection{Construction of a Kriging model in the augmented reliability space}
In this section, the optimization problem in \equaref{eq:005} is solved while each performance function $\mathcal{M}_k$ is replaced by a Kriging model $\widehat{\mathcal{M}}_k$ as introduced above. This simply means that the performance functions for now on read:
\begin{equation}
\widehat{\mathfrak{g}}_k \prt{\ve{x},\ve{z}} = \bar{\mathfrak{g}}_k - \widehat{\mathcal{M}}_k \prt{\ve{x},\ve{z}}, \qquad k = \acc{1,\ldots, n_h}.
\end{equation} 

Computing the quantile with respect to this surrogate model becomes an extremely cheap operation. The expensive part is the initial building of the surrogate model which requires to set and evaluate a design of experiments. Given the possible number of iterations before convergence is achieved, building one surrogate model for each reliability analysis (\ie the quantile computation for a given design in our case) would be quite cumbersome. Instead we advocate for the use of a \emph{single} Kriging model as already proposed in other contributions. The idea is to build the surrogate model in a unique space that embeds both the design and random variables. In \citet{Kharmanda2002}, this space is called the \emph{hybrid design space} and is defined as the tensor product between the design and random variables. This needlessly increases the dimension of the space where the surrogate model is built and may be problematic when it comes to space-filling design of experiments. From another perspective, \citet{Au2005} efficiently computes the failure probability in the so-called \emph{augmented reliability problem} for a given design considering a space where the design variables are artificially considered as random. \citet{Taflanidis2008} use this augmented reliability problem to construct a stochastic optimization problem. Eventually, \citet{Dubourg2011,DubourgThesis} propose an augmented reliability space following the ideas in the two above contributions. In these works, the size of the augmented reliability space remains equal to that of the original reliability problem. This is because they consider that the uncertainty in the random variables is simply augmented by the choice of the design points. In practice, the design and environmental variables are treated separately in the so-called \emph{confidence regions} which span a sufficiently large space such that any point sampled during the analysis is extremely likely to fall within the space of definition of the surrogate model. For the design variables, this region is hyper-rectangular and consists of the design space with extended bounds. The confidence region for the environmental variables is a hypersphere in the standard normal space (\ie, after transforming these variables into standard Gaussian ones) with a sufficiently large radius to account for extreme realizations of the random variables. The augmented space is henceforth considered as the tensor product between these two spaces.

In this paper, we propose an augmented space which is quite close to that defined in \citet{Dubourg2011}. We indeed treat separately the design and environmental variables. However, in our case, the environmental variables are not defined in a hypersphere. The reasons for this are twofold. First, due to its very formulation, the reliability analysis we perform does not need any mapping to the standard Gaussian space. Second, the non-linear mapping from the unit hypersphere (where the space-filling design of experiments is sampled) to the physical space may add complexity and non linearity to the function that is eventually surrogated. To avoid this, we rather consider a hypercube. Since the surrogate models are built in the unit hypercube, the mapping to the physical space is simply linear.

The augmented space is therefore the tensor product between two hyperrectangular confidence regions $\mathbb{X} \times \mathbb{Z}$, where $\mathbb{X}$ refers to the design variables and $\mathbb{Z}$ to the environmental parameters. The former is defined by:
\begin{equation}\label{eq:014}
\mathbb{X} = \prod_{i = 1}^{s_d} \bra{q^-_{d_i},q^+_{d_i}},
\end{equation} 
where $s_d$ is the number of design variables and $q^-_{d_i}$ and $q^+_{d_i}$ are respectively quantiles associated to the lower and upper bounds of the design variables. They are defined in such a way that the confidence region spans a space sufficiently large to contain with high probability (\eg $99 - 99.9 \%$ in the application) all realizations of $\ve{X}|\ve{d}$ sampled during the optimization procedure. They read as follows:
\begin{equation}\label{eq:015}
\begin{split}
q_{d_i}^{-} & = F^{-1}_{X_i|d_i} \prt{\alpha_{d_i}/2} \\
q_{d_i}^{+} & = F^{-1}_{X_i|d_i} \prt{1 - \alpha_{d_i}/2},
\end{split}
\end{equation}
where $X_i$ follows the marginal distribution $f_{X_i|d_i}$, $F^{-1}_{X_i|d_i}$ is the associated inverse CDF, $d_i^-$ and $d_i^+$ are respectively the lower and upper bounds of the design variable $d_i$, and $\alpha_{d_i}$ is the probability of sampling outside the augmented space. In applications we select $\alpha_{d_i} = 2.7 \; 10^{-3}$ for each variable, which corresponds to $\mu \pm 3 \sigma$ for a Gaussian variable.

In the same fashion, the confidence region for the environmental variables is defined by:
\begin{equation}\label{eq:016}
\mathbb{Z} = \prod_{j = 1}^{s_z} \bra{q^-_{z_j},q^+_{z_j}},
\end{equation}
where $s_z$ is the number of environmental variables and the bounding quantiles are defined by:
\begin{equation}\label{eq:409}
\begin{split}
q_{z_j}^- & = F^{-1}_{Z_j} \prt{\alpha_{z_j}/2}, \\
q_{z_j}^+ & = F^{-1}_{Z_j} \prt{1 - \alpha_{z_j}/2},
\end{split}
\end{equation}
where $Z_j$ follows the marginal distribution $f_{Z_j}$ whose inverse CDF is $F^{-1}_{Z_j}$ and $\alpha_{z_j}$ is the probability of sampling outside the augmented space in the direction of $Z_j$, again in the order of $10^{-3}$ in applications.

\begin{figure}[!ht]
	\begin{center}
		\includegraphics[width=0.65\textwidth]{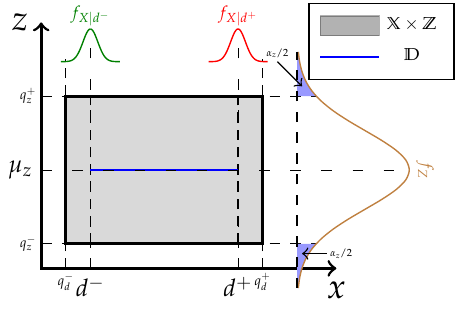}
		\caption{Illustration of the augmented space for a two-dimensional problem with both random design and environmental variables.} %
		\label{fig:Aug_Spa}%
	\end{center}
\end{figure}

To illustrate the augmented space defined in this paper, we consider a problem where the design space is one-dimensional: $\mathbb{D} = \bra{d^-,d^+}$. For the RBDO problem, the design variable is supposed random with distribution $d \sim \mathcal{N}\prt{d,\sigma_d^2}$. We also assume that the RBDO problem features a unique environmental random variable defined by $Z \sim \mathcal{N}\prt{\mu_z, \sigma_z^2}$. An augmented space associated to this problem is shown in \figref{fig:Aug_Spa}. The design space is the blue line and the augmented space is the gray area. The distributions of the design and environmental variables are also plotted.

Note that the proposed framework naturally encompasses the two following cases:
\begin{itemize}
	\item When the analysts disregard uncertainties in the design parameters (\eg, ignore manufacturing tolerances), $q_{d^-}$ and $q_{d^+}$ are simply set to $d^-$ and $d^+$ respectively.
	\item When no environmental variables are considered, the augmented space reduces to $\mathbb{X}$.
\end{itemize}
Once the augmented space is defined for a specific problem, one may build a single \emph{global} surrogate model. This may be achieved by a space-filling design of experiments, \ie by using uniformly distributed samples so as to cover the entire space. The built Kriging model may henceforth be used for any reliability analysis during the RBDO. This one-shot approach is theoretically possible  but would require the Kriging model to be accurate in the entire space. However, during the optimization only a subset of the space is actually of interest, \ie regions in the vicinity of the limit-state surface and those where the objective function decreases. These two issues can be dealt with using so-called \emph{adaptive design of experiments}. We propose in this paper a two-stage enrichment scheme where each stage is geared toward achieving one of the two above goals.

\subsection{Adaptive design of experiments: A short literature review}
Adaptive design of experiments have been developed from the premise that only a limited region of the space is of interest to the designer during an optimization analysis. Thus, instead of densely filling the space so as to have an evenly accurate model in the entire space, the optimization starts with a not so accurate model built upon a scarcely sampled design of experiments. Enrichment is then made so as to improve the surrogate accuracy in regions that matter. \citet{Jones1998} proposed the \emph{efficient global optimization} (EGO) scheme relying on an \emph{expected improvement} function which focuses on sequentially updating a Kriging model so as to converge to a global minimum. From the same idea, numerous authors have proposed infill sampling criteria to achieve the same goal. As for RBDO, emphasis is rather put on the vicinity of the limit-state surface in order to accurately estimate failure probabilities. A first family of infill sampling criteria comes from EGO techniques as they are mere adaptation, \eg adjusted expected improvement \citep{schonlau1998}, expected violation \citep{Audet2000} or expected improvement for contour approximation \citep{Ranjan2008}. On the other hand, \citet{Bichon2008} introduced a so-called \emph{efficient global reliability analysis} (EGRA) where an \emph{expected feasibility function} is used to improve the surrogate model in the vicinity of the limit-state surface. Similarly, \citet{DubourgThesis} used in his PhD thesis work the margin probability function. In the present work, we will focus on the \emph{deviation number} developed by \citep{Echard2011} for their Active Kriging Monte Carlo simulation technique (AK-MCS). In AK-MCS, some candidates to enrichment are considered among a Monte Carlo set. The point that is most likely to improve the Kriging model is defined as the one that minimizes the following $U$-function:
\begin{equation}\label{eq:018}
U\prt{\ve{x}} = \frac{\abs{\bar{\mathfrak{g}} - \mu_{\widehat{\mathcal{M}}} \prt{\ve{x}} } }{\sigma_{\widehat{\mathcal{M}}} \prt{\ve{x}}}.
\end{equation}
In practice, points that tend to minimize this function are those which are close to the constraint threshold \ie $\mu_{\widehat{\mathcal{M}}}\prt{\ve{x}} \rightarrow \bar{\mathfrak{g}}$ (otherwise put, $ \widehat{\mathfrak{g}}\prt{\ve{x}} \rightarrow 0$), or those for which the Kriging variance is high ($\sigma_{\widehat{\mathcal{M}}} \prt{\ve{x}} \rightarrow \infty$), thus implying that the Kriging model may lack of accuracy there because of the DoE scarcity.

In the sequel, we adapt this function for contour estimation with respect to a quantile which is referred to as the global stage of enrichment.

\subsection{Proposed global stage of enrichment} \label{SEC:glo_enr}
This first stage of enrichment is aimed at revealing regions of the space where the constraints, as defined in terms of quantiles, are likely to be violated. We call it \emph{global} as this enrichment spans the entire augmented space just as the AK-MCS defined above. There is however one difference in our setting. In contrast to AK-MCS, the constraint is defined with respect to $\ve{d}$ in the design space but the Kriging model is built in the augmented space. The idea with the proposed approach is to find the pair of points in the augmented space that most likely leads to an improvement of the quantile estimation in the design space. The following steps are completed to achieve this task:
\begin{enumerate}
	\item Sample candidates for enrichment in the design space:
	\begin{equation}\label{eq:019}
	\mathfrak{C} = \acc{\ve{d}^{(1)}, \ve{d}^{(2)}, \ldots, \ve{d}^{(m)} }	
	\end{equation}
	\item For each design $\ve{d}^{(i)}, i = \acc{1, \ldots, m}$:
	\begin{enumerate}
		\item Sample the Monte Carlo set required to compute the quantile:
		\begin{equation}
		\mathfrak{C}_q^{(i)} = \acc{\prt{\ve{x}_j, \ve{z}_j}, j = 1, \ldots, N}
		\end{equation}
		\item Compute the associated quantile $\widehat{q}_{\alpha} \prt{\ve{d}^{(i)}}$
		\item Identify the point in the augmented space that is associated to the quantile, \ie
		\begin{equation}\label{eq:021}
		\prt{\ve{x}_\alpha^{(i)},\ve{z}_\alpha^{(i)}} = \acc{ \prt{\ve{x},\ve{z}} \in \mathfrak{C}_q^{(i)} \; : \, \widehat{q}_\alpha \prt{\ve{d}^{(i)}} = \mu_{\widehat{\mathcal{M}}} \prt{\ve{x},\ve{z}}}
		\end{equation}
		\item Compute the modified deviation number:
		\begin{equation}\label{eq:022}
		\mathcal{U}\prt{\ve{d}^{(i)}}   \equiv U \prt{\ve{x}_\alpha^{(i)},\ve{z}_\alpha^{(i)}} = \frac{\abs{\bar{\mathfrak{g}} - \mu_{\widehat{\mathcal{M}}} \prt{\ve{x}_\alpha^{(i)},\ve{z}_\alpha^{(i)}}}} {\sigma_{\widehat{\mathcal{M}}} \prt{\ve{x}_\alpha^{(i)},\ve{z}_\alpha^{(i)}}}
		\end{equation}
	\end{enumerate}
	\item The next best point to add to the design of experiments is therefore defined as:
	\begin{equation}\label{eq:462}
	\prt{\ve{x}_{\text{next}},\ve{z}_{\text{next}}} = \arg \min_{\prt{\ve{x}_{\alpha},\ve{z}_{\alpha}} \in \mathfrak{C}_\alpha} \mathcal{U} \prt{\ve{d}},
	\end{equation}
	where $\mathfrak{C}_\alpha = \acc{\prt{\ve{x}_\alpha^{(i)},\ve{z}_\alpha^{(i)}}, i = 1, \ldots, m}$.
\end{enumerate}

\begin{figure}[!ht]
  \centering \subfloat[Augmented space (left) and design space (right):
  iteration $\#
  1$]{\includegraphics[width=0.49\textwidth]{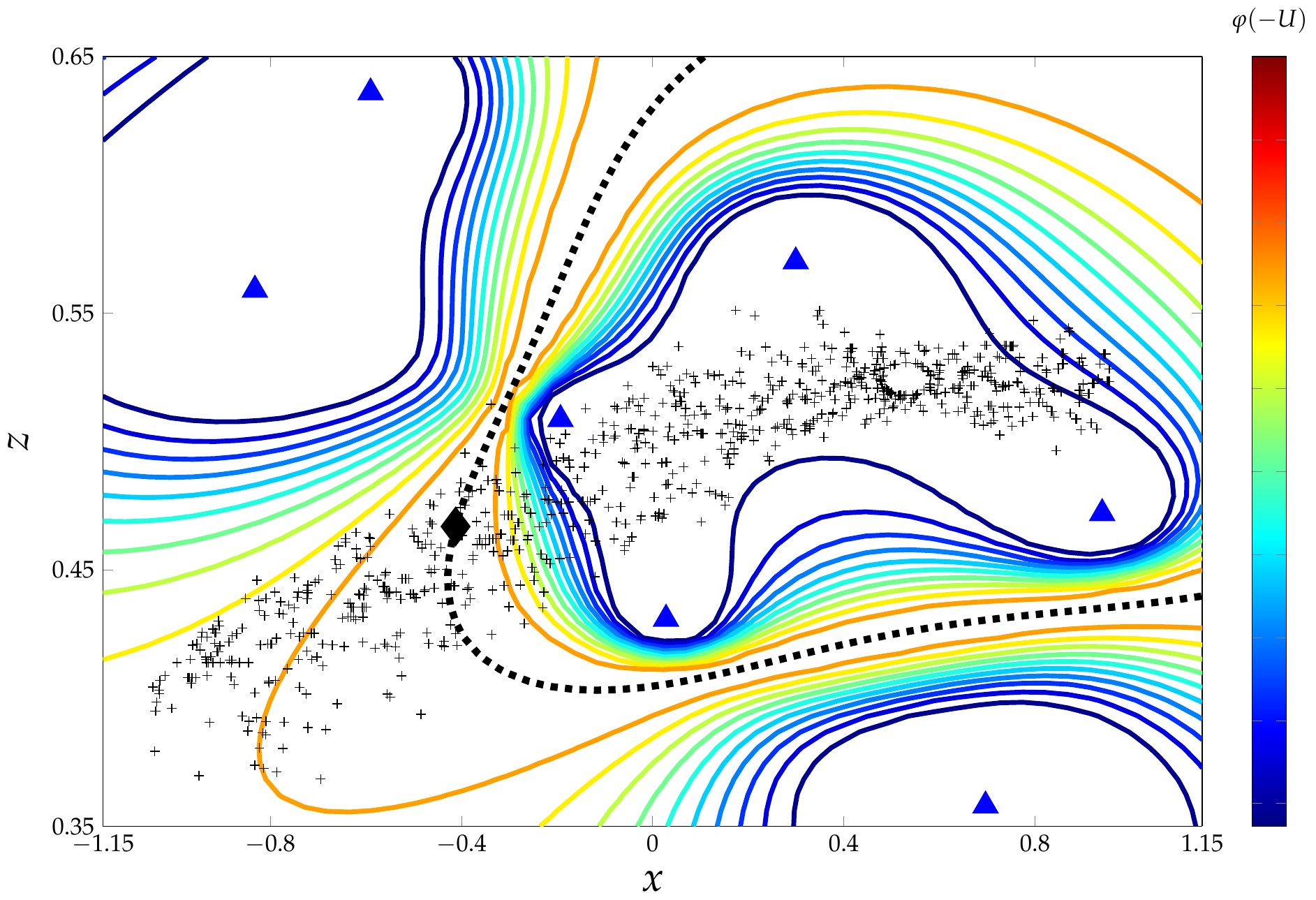} \hfill
    \includegraphics[width=0.45\textwidth]{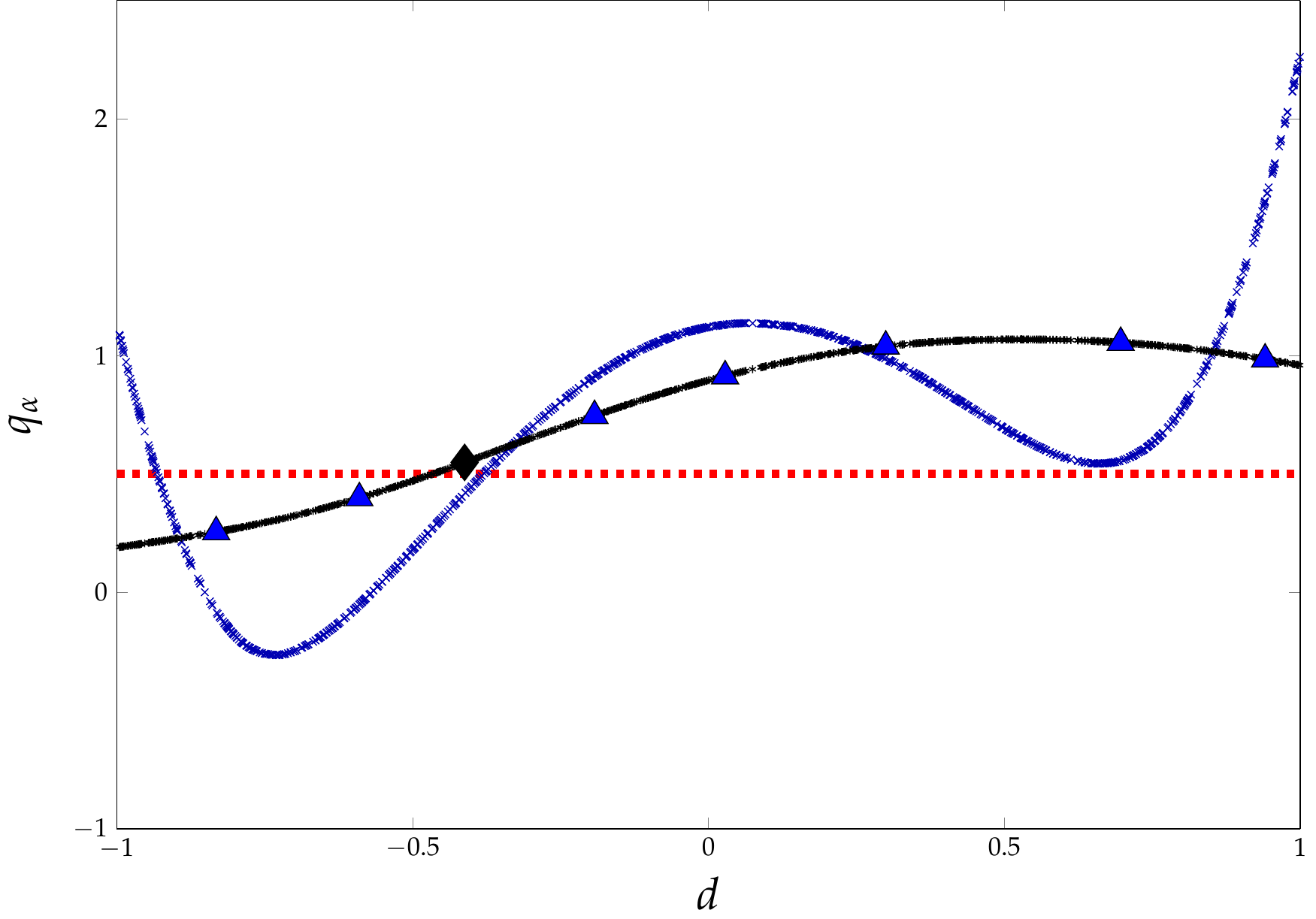}\label{fig:qea}}%
  \\
  \subfloat[Augmented space (left) and design space (right): iteration
  $\# 10$]{\includegraphics[width=0.49\textwidth]{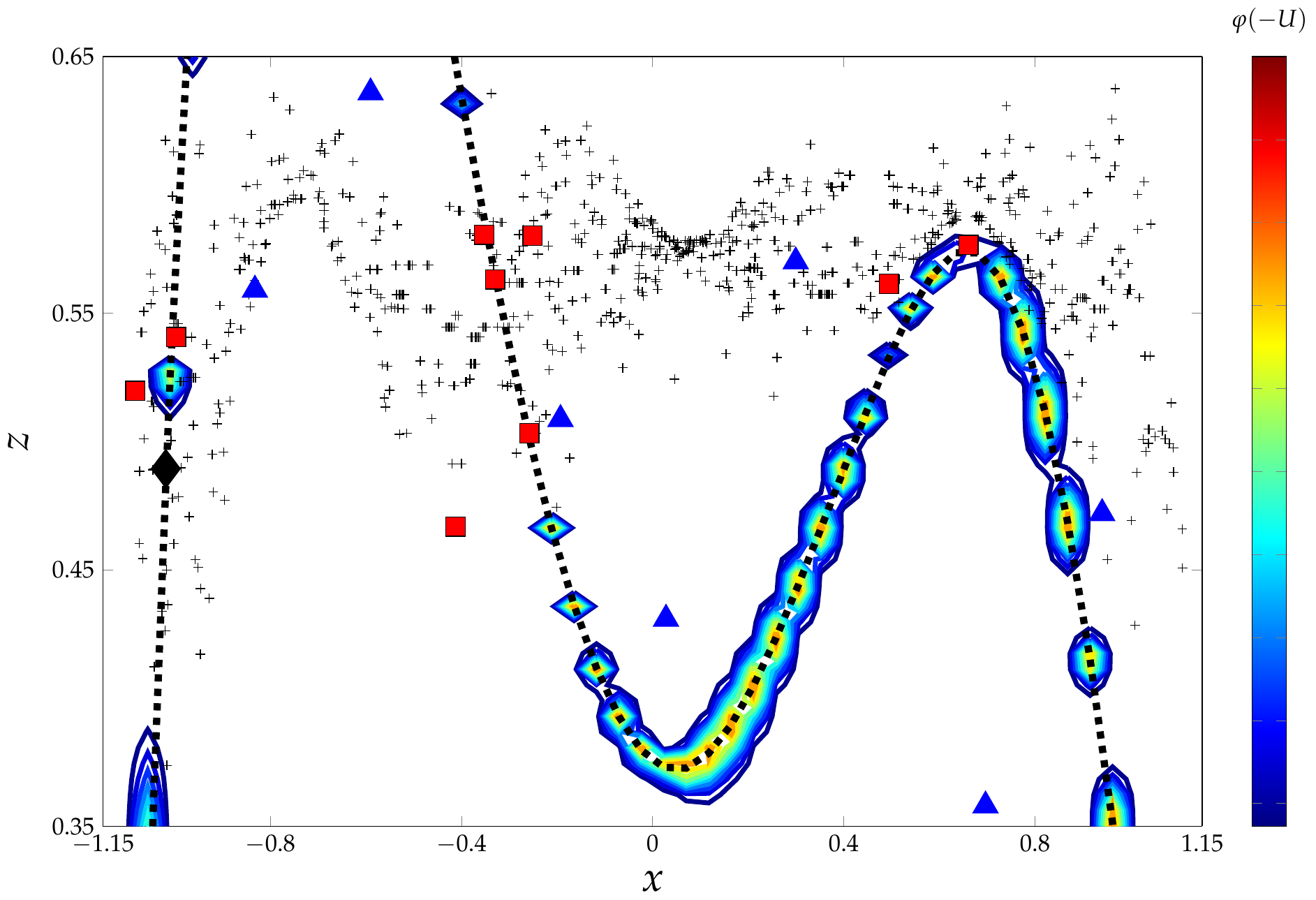}
		\includegraphics[width=0.45\textwidth]{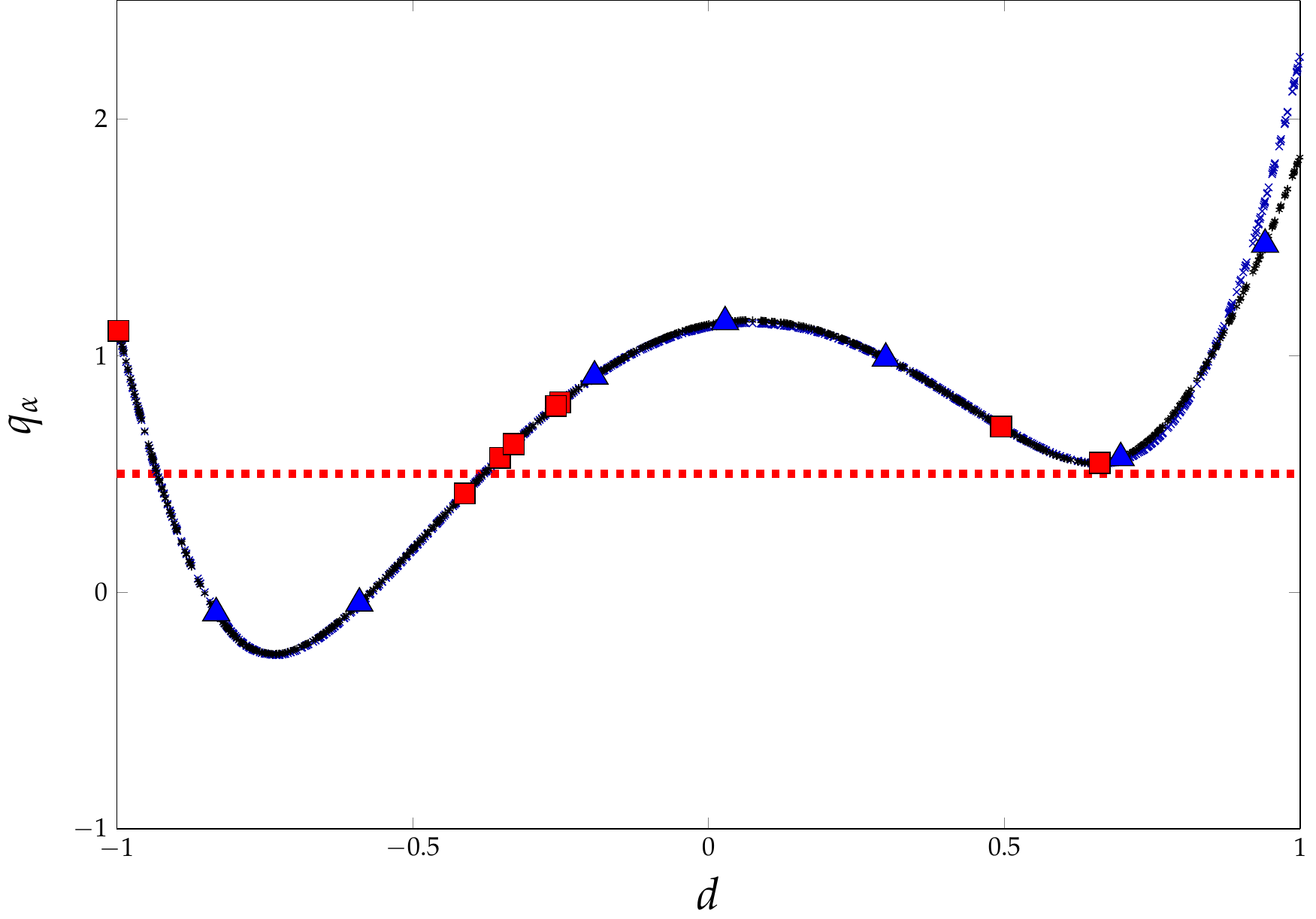}\label{fig:qed}}%
	\caption{Enrichment with the mathematical function. In the left panel, the augmented space with contours of the enrichment functions and the set $\mathfrak{C}_\alpha$  shown as small crosses. In the right panel, the true and estimated quantiles are shown (resp. blue and black curve). Triangles and squares respectively stand for initial and enrichment points.}%
	\label{fig:quant_enrich}%
\end{figure}

To illustrate this enrichment scheme, let us consider the mathematical function from \citet{Janusevskis2013} which reads:
\begin{equation}\label{eq:293}
\mathcal{M}\prt{d,z} = \prt{\frac{1}{3}z^4 - 2.1 z^2 + 4}z^2 + dz + 4d^2 \prt{d^2 - 1},
\end{equation}
where $d \in \bra{-1,1}$. It is considered as a performance function for an RBDO problem where the constraint threshold is set to $\bar{\mathfrak{g}} = 0.5$. The probabilistic model consists of the random design variable $X \sim \mathcal{N}\prt{d,0.05^2}$ and the random environmental variable $Z \sim \mathcal{N}\prt{0.5, 0.05^2}$. \figref{fig:quant_enrich} shows the various iterations of the enrichment procedure. The left panel shows the augmented space with contour of $\varphi(-U)$, where $\varphi$ is the standard Gaussian PDF ($U$ is conveniently mapped for proper illustration). The contour $\widehat{\mathfrak{g}}\prt{\ve{x}, \ve{z}} = 0$ is plotted as black dotted line and the small black crosses form the set $\mathfrak{C}_\alpha$. In the right panel, the true and estimated quantiles ($\mathfrak{q}_\alpha$ and $\widehat{\mathfrak{q}}_\alpha$) are respectively plotted in blue and black lines. The threshold $\bar{\mathfrak{g}} = 0.5$ is represented by the red dotted line. The blue triangles are the initial DoE. As enrichment is carried out, the red squares are added to the DoE. At each iteration, the best next point corresponds to the black diamond. From this example, we can see that the points added in the augmented space actually corresponds to those where $\widehat{\mathfrak{q}}_\alpha\prt{d} \rightarrow  \bar{\mathfrak{g}}$.

In this example, the enrichment was stopped when $\min_{d \in \mathfrak{C}} \mathcal{U} \geq 2$. This criterion actually means that there is only $5 \%$ of chance of mistaking a safe design for a failed one (and vice-versa) w.r.t. all the points in $\mathfrak{C}$. This is quite a conservative stopping criterion. We do not need such an accuracy in the entire design space. Since the next step is optimization, we may go further in reducing the computational budget by saving model evaluations to regions that actually improve the objective function. For this reason we propose a second stage of enrichment as explained in the sequel.

\subsection{Local stage of enrichment}

In order to further reduce the number of calls to the original computational model, we stop the first and global stage of enrichment earlier and proceed to a local enrichment which is coupled with optimization. In fact, the idea is to have a roughly accurate surrogate model that reveals the different regions of the space where the approximated limit-state function is close to zero. To this end, we may relax the criterion $\min \mathcal{U} \geq 2$ to a certain proportion of the enrichment candidates rather than all of them. The criterion may therefore be written as:
\begin{equation}\label{eq:025}
\eta = \text{Card}\prt{\mathfrak{C}_2} / \text{Card}\prt{\mathfrak{C}} \leq \bar{\eta}.
\end{equation}
where $\mathfrak{C}_2 = \acc{\ve{d} \in \mathfrak{C} \; : \; \mathcal{U}\prt{\ve{d}} \leq 2}$. Note that the original criterion corresponds to $\bar{\eta} = 0$. We may consider a relaxed criterion by setting $\bar{\eta} = 0.30$ for instance.

Assuming that the first stage of enrichment has been stopped with respect to the above criterion, there is residual epistemic uncertainty to the Kriging model. This uncertainty can be monitored during optimization and dealt with by updating the Kriging model only when necessary. To achieve this goal, we may consider a local accuracy measure associated to the quantile estimates, as they ultimately define the constraints of interest.

Following the idea in \citet{Dubourg2011} where bounds on failure probabilities were developed, we define the following lower and upper bounds, $\widehat{\mathfrak{q}}_\alpha^-$ and $\widehat{\mathfrak{q}}_\alpha^+$, which are quantiles computed with respect to $\mu_{\widehat{\mathcal{M}}} - 2 \sigma_{\widehat{\mathcal{M}}}$ and $\mu_{\widehat{\mathcal{M}}} + 2 \sigma_{\widehat{\mathcal{M}}}$. Since the standard deviation is positive the following relationship holds:
\begin{equation}
\widehat{\mathfrak{q}}_\alpha^-\prt{\ve{d}} \leq \widehat{\mathfrak{q}}_\alpha\prt{\ve{d}} \leq \widehat{\mathfrak{q}}_\alpha^+\prt{\ve{d}} \qquad \text{for any } \ve{d} \in \mathbb{D}.
\end{equation}
The spread of this interval is a good measure of the local Kriging accuracy for the quantile estimation. The following local accuracy criterion may henceforth be derived:
\begin{equation}
\eta_q\prt{\ve{d}} = \frac{\widehat{q}_\alpha^+\prt{\ve{d}} - \widehat{q}_\alpha^-\prt{\ve{d}}}{\bar{\mathfrak{g}}} \leq \bar{\eta}_q,
\end{equation}
where $\bar{\eta}_q$ is a pre-defined threshold. In the case where $\bar{\mathfrak{g}} = 0$, we may replace the denominator by $\sqrt{\Var{\widehat{\mathcal{Y}}_{MCS}}}$, where $\Var{\widehat{\mathcal{Y}}_{MCS}}$ is the variance of Kriging prediction over a large Monte Carlo set sampled in the augmented space.

The surrogate model is considered to be accurate enough for the quantile
estimation at the design $\ve{d}^{(i)}$ if this relationship holds. If
in contrast $\eta_q > \bar{\eta}_q$, then a local enrichment is made. To
this end, candidates for enrichment are selected among the Monte Carlo
set $\mathfrak{C}_q^{(i)}$. The following deviation number is computed
on this set \citep{Schobi2014,SchoebiASCE2016}:
\begin{equation}\label{eq:028}
\mathfrak{U}\prt{\ve{x},\ve{z}} = \frac{\abs{\widehat{\mathfrak{q}}_\alpha\prt{\ve{d}^{\prt{i}}} - \mu_{\widehat{\mathcal{M}}}\prt{\ve{x},\ve{z}}}} {\sigma_{\widehat{\mathcal{M}}}\prt{\ve{x},\ve{z}}}.
\end{equation}
The best next point is the one that minimizes this function. This point corresponds to a certain $\prt{\ve{x}_\alpha^{(i)},\ve{z}_\alpha^{(i)}}$ from \equaref{eq:021}. By iteratively adding points in this fashion, it is expected that the quantile will be more and more accurately estimated.

\subsection{Implementation of the proposed procedure}

We now consider the implementation of the whole procedure. Prior to that, let us specify two additional points that most often characterize the actual problems we intend to address, \ie the possibility of adding many points per iterations and the presence of multiple constraints. The first point may be interesting when one has computational resources that allow for distributed computations. It may also be argued that there is not one single point that is likely to improve the surrogate model but many points located in disjoint regions. In such a case, multiple enrichment points allow us to reach them simultaneously. In order to add $K$ points in the DoE, we consider a weighted $K$-means clustering of the  candidates for enrichment, where each point is weighted by $\varphi(-U)$. This way, regions with small values of $U$ are favored. Finally, $K$ clusters centers are chosen as the next points to add in the DoE.

As for the case of multiple constraints, many techniques exist. We may, for instance, rank the constraints and enrich sequentially starting with the most important one. This is not an optimal scheme. \citet{Fauriat2014} proposed a composite criterion which focuses on the most violated constraints. However, the notion of "most violated" is not adequate when the constraints are defined on completely different scales. In this work, we thus consider a composite criterion where, for each enrichment candidate, the constraint with minimum value of $U$ is taken, that is:
\begin{equation}\label{eq:029}
\begin{split}
\mathcal{U}_{comp} \prt{\ve{x},\ve{z}} = & \min_{l \in \acc{1, \ldots, n_h}} \mathcal{U}_{l} \prt{\ve{x},\ve{z}}= \frac{ \abs{\bar{\mathfrak{g}}_l - \mu_{\widehat{\mathcal{M}}_l}\prt{\ve{x} ,\ve{z}}}}{\sigma_{\widehat{\mathcal{M}}_l}\prt{\ve{x},\ve{z}}}, \\
\mathfrak{U}_{comp} \prt{\ve{x},\ve{z}} = & \min_{l \in \acc{1, \ldots, n_h}} \mathfrak{U}_l\prt{\ve{x},\ve{z}} = \frac{\abs{\widehat{\mathfrak{q}}_{\alpha_l}\prt{\ve{d}^{\prt{i}}} - \mu_{\widehat{\mathcal{M}}_l}\prt{\ve{x},\ve{z}}}} {\sigma_{\widehat{\mathcal{M}}_l}\prt{\ve{x},\ve{z}}}.
\end{split}
\end{equation}

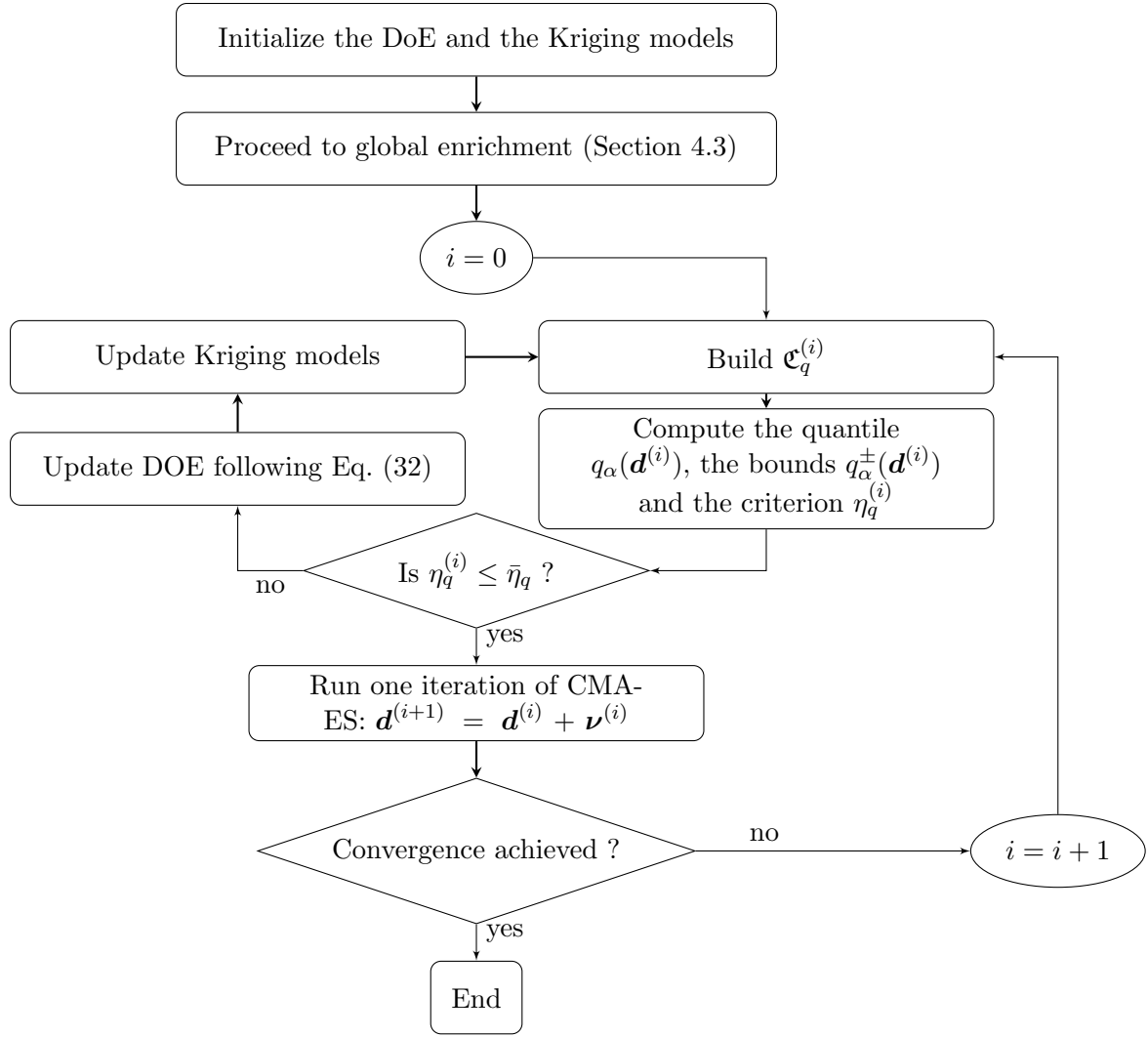
\begin{figure}[!ht]
	\begin{tikzpicture}[node distance=1.5cm,auto,scale=0.99,transform shape]
	
	\node(init)[init] {Initialize the DoE and the Kriging models};
	
	\node(glo_enr)[glo_enr, below of = init]{Proceed to global enrichment (Section \ref{SEC:glo_enr})};
	
	\node (i0) [counter, below of =glo_enr] {$i=0$};
	
	\node(Cq)[optimization,below right = 0.5 cm and 0.3 cm of i0]{Build $\mathfrak{C}_q^{(i)}$};
	
	\node (comp)[optimization, below = 0.2 cm of Cq]{Compute the quantile $q_\alpha(\ve{d}^{(i)})$, the bounds $q^\pm_\alpha(\ve{d}^{(i)})$ and the criterion $\eta_q^{(i)}$} ;
	
	\node (eta) [decision, below = 3 cm of i0,aspect=3] {Is $\eta_q^{(i)} \leq \bar{\eta}_q$ ?};
	
	\node(udoe)[optimization, left = 1cm of comp]{Update DOE following \equaref{eq:029}};
	
	\node (meta)[optimization, left = 1cm of Cq ]{Update Kriging models};
	
	\node (ddi)[optimization, below = 0.5 cm of eta ]{Run one iteration of CMA-ES: $\ve{d}^{(i+1)} = \ve{d}^{(i)} + \ve{\nu}^{(i)}$};
	
	\node (conv) [decision, below = 0.5 cm of ddi ,aspect=3] {Convergence achieved ?};

	\node (fin) [Fin, below = 0.5 cm of conv, node distance = 1.5cm] {End};
	\node (ii1) [counter, right of = conv, node distance = 8cm] {$i=i+1$};

	\draw [arrow] (init) -- (glo_enr);
	\draw [arrow] (glo_enr) -- (i0);
	\path [line] (i0) -|  (Cq);
	\draw [arrow] (Cq) -- (comp);
	\path [line] (comp) |-  (eta);
	\path [line] (eta) -- node[near start] {yes}  (ddi);
	\path [line] (eta) -|  node[near start] {no} (udoe);
	\draw [arrow] (udoe) -- (meta);
	\draw [arrow] (meta) -- (Cq);
	\path [line] (conv) --  node[near start] {no} (ii1);
	\path [line] (ii1) |-  (Cq);
	\draw [arrow] (ddi) -- (conv);
	\path [line] (conv) -- node[near start] {yes}  (fin);
	\end{tikzpicture}
	\caption{Flowchart of the optimization procedure with the two stages of enrichment.}
	\label{fig:flowchart}
\end{figure}

Considering all these developments, the pseudo-code in Algorithm~\ref{alg:qRBDO} summarizes the proposed procedure. Here we consider that the first stage of enrichment has been already performed. The selected optimization algorithm is the $(1+1)$-CMA-ES (Covariance matrix adaptation - evolution strategy) for constrained problems \citep{Arnold2012}. This is a stochastic global search algorithm which relies on multivariate normal distributions to search candidates with increased fitness as iterations grows. It also accounts for constraints by decreasing the likelihood to sample in the direction of previously unfeasible sampled points. Such a global search algorithm is quite convenient for the proposed procedure since only one parent generates one offspring, thus allowing us to check the quantile accuracy for the offspring before moving on. The entire procedure is illustrated in \figref{fig:flowchart}.

\begin{spacing}{0.8}
\begin{algorithm}
	\caption{Quantile and adaptive Kriging optimization procedure}
	\begin{algorithmic}[1]
		\Require{}
		\Statex DoE after the first stage of enrichment $\mathcal{D}$
		\Statex Kriging models $\acc{\widehat{\mathcal{M}_l}, l=1, \ldots, n_h}$ based on the DoE $\mathcal{D}$
		\Statex Target failure probability $\acc{\bar{P}_{f_l} = 1 - \alpha_l, l = 1, \ldots, n_h}$
		\Statex Initial design for optimization $\ve{d}^{(0)}$
		\Statex Number of simultaneous enrichment points $K$ \Comment{\color{BlueViolet} {\scriptsize{\eg $K = 3$}} \color{black}}
		\Statex Constraint and quantile accuracy thresholds $\bar{\mathfrak{g}}$ and $\bar{\eta}_q$ \Comment{\color{BlueViolet} {\scriptsize{\eg $\bar{\eta}_q = 0.1$}} \color{black}}
		\Statex Size of the Monte Carlo set $\mathfrak{C}_q$  $N$ \Comment{\color{BlueViolet} {\scriptsize{\eg $N = 10,000$}} \color{black}}
		\Statex \hrulefill
		\State $i = 0$; NotConverged $= \texttt{true}$,
		\While{NotConverged $= \texttt{true}$}
		\State Draw samples $\mathfrak{C}_q^{(i)} = \acc{\prt{\ve{x}_1,\ve{z}_1}, \ldots \prt{\ve{x}_N,\ve{z}_N}}$ in the augmented space where $\ve{X} \sim f_{\ve{X}|\ve{d}^{(i)}}$ and $\ve{Z} \sim f_{\ve{Z}}$ 
		\For{$l = 1 \textrm{ to } n_h$}
		\For{$j = 1 \textrm{ to } N$}
		\State $\widehat{y}_j = \mu_{\widehat{\mathcal{M}}_l} \prt{\ve{x}_j,\ve{z}_j}$
		\State $\widehat{y}_j^{-} = \mu_{\widehat{\mathcal{M}}_l} \prt{\ve{x}_j,\ve{z}_j} -  2 \sigma_{\widehat{\mathcal{M}}_l} \prt{\ve{x}_j,\ve{z}_j} $
		\State $\widehat{y}_j^{+} = \mu_{\widehat{\mathcal{M}}_l} \prt{\ve{x}_j,\ve{z}_j} +  2 \sigma_{\widehat{\mathcal{M}}_l} \prt{\ve{x}_j,\ve{z}_j} $
		\EndFor
		\State $q_{\alpha_l}\prt{\ve{d}^{(i)}} = \texttt{quantile}\prt{\acc{\widehat{y}}_{j=1}^{N},\alpha_l}$ \Comment{\color{OliveGreen} {\scriptsize{Estimated quantile}} \color{black}}
		\State $q_{\alpha_l}^{-}\prt{\ve{d}^{(i)}} = \texttt{quantile}\prt{\acc{\widehat{y}^{-}}_{j=1}^{N},\alpha_l}$ \Comment{\color{OliveGreen} {\scriptsize{Lower bound of the quantile}} \color{black}}
		\State $q_{\alpha_l}^{+}\prt{\ve{d}^{(i)}} = \texttt{quantile}\prt{\acc{\widehat{y}^{+}}_{j=1}^{N},\alpha_l}$	\Comment{\color{OliveGreen}  {\scriptsize{Upper bound of the quantile}}\color{black}}
		\EndFor
		\If{$\prt{q_\alpha^{+} - q_\alpha^{-}}/\bar{\mathfrak{g}} > \bar{\eta}_q$}
		\For{$k = 1 \textrm{ to } N$}
		\For{$l = 1 \textrm{ to } n_h$}
		\State $\mathfrak{U}_l\prt{\ve{x}_k,\ve{z}_k} =  \abs{\mu_{\widehat{\mathcal{M}}_l } \prt{\ve{x}_k,\ve{z}_k} - q_{\alpha_l}}/\sigma_{\widehat{\mathcal{M}}_l} \prt{\ve{x}_k,\ve{z}_k}$
		\EndFor
		\State $ \mathfrak{U}_{comp} \prt{\ve{x}_k,\ve{z}_k} = \min_{l \in \acc{1, \ldots, n_h}} \mathfrak{U}_l\prt{\ve{x}_k,\ve{z}_k}$
		\EndFor
		\If{$K == 1$}
		\Comment{\color{OliveGreen} {\scriptsize{The point that minimizes $\mathfrak{U}_{comp}$ is chosen}} \color{black}} \State $\prt{\ve{x}_{\text{next}},\ve{z}_{\text{next}}} = \arg \min \acc{\mathfrak{U}_{comp}\prt{\ve{x}_k,\ve{z}_k}}_{k=1}^{N}$ 
		\Else \Comment{\color{OliveGreen} {\scriptsize{$K$ points are chosen among the $N$ candidates}}\color{black}}
		\State $\prt{\ve{x}_{\text{next}},\ve{z}_{\text{next}}}$ obtained from weighted $K$-means clustering with weight $\varphi\prt{-\mathfrak{U}_{comp}}$ 
		\EndIf
		\EndIf
		\State Update the DoE $\mathcal{D}$ and the Kriging models $\widehat{\mathcal{M}_l}$
		\Let{$\ve{d}^{(i)}$}{$\ve{d}^{(i)} + \nu^{(i)}$} \Comment{\color{OliveGreen} {\scriptsize{Explore the next design point using $(1+1)$-CMA-ES}} \color{black}}
		\Let{$i$}{$i+1$}
		\State Check convergence of the optimization algorithm
		\EndWhile
	\end{algorithmic}
	\label{alg:qRBDO}
\end{algorithm}
\end{spacing}
\newpage
\section{Application examples}
The proposed methodology is now validated with four application examples. The first three are analytical problems whose solutions are available in the literature. The last one is related to the lightweight design of an automotive body structure under crashworthiness constraints. The following settings are common to all the problems. $L_2$-discrepancy-based optimized Latin hypercube is used to generate the initial designs of experiments. Anisotropic Kriging with Mat\'ern 5/2 autocorrelation function and a constant trend is considered as the default surrogate model.

\subsection{Column under compression}
This first example, introduced in \citet{DubourgThesis}, is concerned with a column of rectangular cross-section $b \times h$ submitted to a compressive load $F_{ser}$. The aim is to minimize the cross-sectional area while avoiding buckling. Buckling may occur here if the service load is higher than critical Euler force which reads:
\begin{equation}\label{eq:030}
F_{cr} = \frac{\pi^2 E I}{L^2},
\end{equation}
where $L$ is the length of the column, $E$ is the Young's modulus of its constitutive material and $I= bh^3/12$ ($b > h$) is the column area moment of inertia.

The deterministic optimization problem then reads:
\begin{equation}\label{eq:031}
\begin{split}
& \ve{d}^\ast = \arg \min_{\ve{d} \in \bra{150,350}^2} bh  \quad \text{subject to: }
\left\{ \begin{array}{ll}
\displaystyle{\mathfrak{f}\prt{\ve{d}} = h - b  \leq 0,} \\
\displaystyle{\mathfrak{g}\prt{\ve{d},\ve{z}} =F_{ser} - k \frac{\pi^2 E b h^3}{12 L^2},}
\end{array} \right.
\end{split}
\end{equation} 
where $k$ is a parameter which accounts for model uncertainty in the Euler force (\ie it represents the effect of imperfections in the beam geometry and may be viewed as a model correction factor with respect to the ideal Euler force) and $\ve{z} = \acc{k,E,L}^T$ is the vector of environmental variables. Uncertainties are considered by introducing the probabilistic model as described in \tabref{tab:001}. With all parameters being lognormally distributed, an analytical solution can be derived \citep{DubourgThesis}:
\begin{equation}\label{eq:032}
b^\ast = h^\ast = \frac{12 F_{ser}}{\pi^2 \exp \prt{ \lambda_k + \lambda_E - 2 \lambda_L + \Phi^{-1}\prt{P_f} \sqrt{\zeta_k^2 + \zeta_E^2 + 4 \zeta_L^2}}},
\end{equation}
where $\zeta_\bullet = \sqrt{\ln \prt{1+\delta_\bullet^2}}$ and $\lambda_\bullet = \ln\prt{\mu_\bullet} - \frac{1}{2} \zeta_\bullet^2$ are respectively the scale and location parameters of the lognormal distribution. By setting the target probability of failure to $5 \%$, \ie $\alpha = 0.95$, the analytical solution, $b^\ast = h^\ast = 238.45$ mm.

\begin{table}
	\centering
	\caption{Probabilistic model for the column under compression example.}
	\label{tab:001}
	\begin{tabular}{lccc}
		\hline
		{Parameter}    & {Distribution}  & {Mean ($\mu$)}  & {COV ($ \delta \%$)} \\ \hline
		{$k$}		      & {Lognormal}		& {$0.6$}	   & {$10$}	   \\
		{$E$ (MPA)}	  & {Lognormal}	  & {$10,000$}  & {$5$}   \\
		{$L$ (mm)}		  &{Lognormal}	  & {$3,000$}	 & {$1$}    \\
		{$F_{ser}$ (N)}		  &{$-$}	  & {$1.4622 \times 10^6$}	 & {$-$}    \\
		\hline
	\end{tabular}
\end{table}

\begin{figure}[!ht]
	\begin{center}
          \subfloat[Points sampled during CMA-ES in the design
          space]{\includegraphics[width=0.49\textwidth]{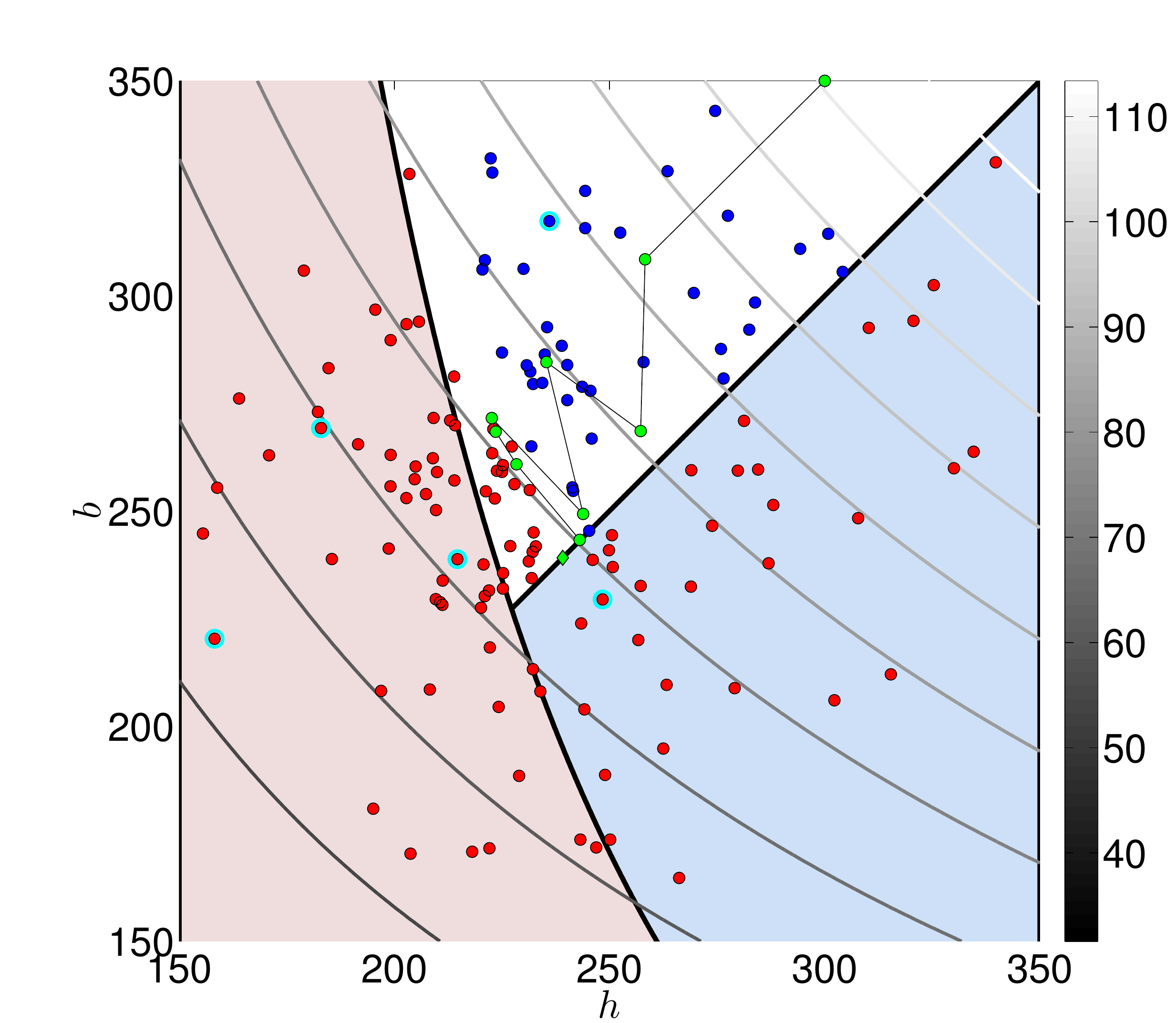}\label{fig:Euler_cmaes}}%
          \hfill%
          \subfloat[Quantile accuracy criterion during
          optimization]{\includegraphics[width=0.49\textwidth]{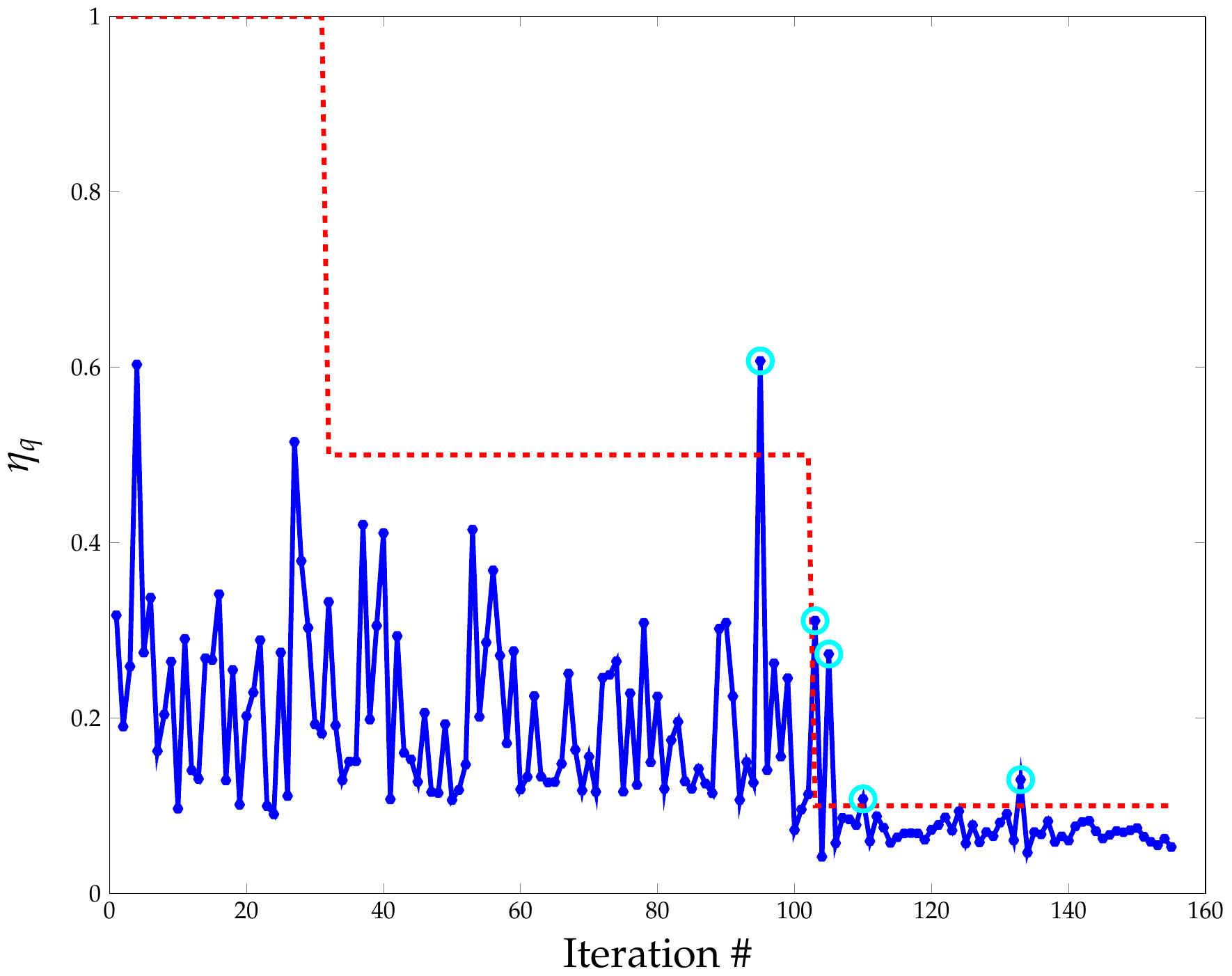}\label{fig:Euler_conv}}%
		\caption{Convergence of the column under compression problem. The left panel shows the evolution of CMA-ES in the design space. The blue and red dots fall respectively in the feasible and unfeasible sets. The green ones are the successive best sample points. In the right panel, the evolution of the local accuracy criterion is shown with respect to the number of iterations. In the two figures, the points corresponding to enrichment have been circled in cyan.} %
		\label{fig:Euler_appl}%
	\end{center}
\end{figure}

To apply the methodology on this five-dimensional problem, we start with a scarce initial design of $10$ points and set the global accuracy threshold in \equaref{eq:025} to $\bar{\eta} = 0.15$. Only two enrichment points are necessary to reach the required global accuracy. We then start the optimization by setting a simulated-annealing-like threshold $\bar{\eta}_q$ with three levels which are respectively $1$, $0.5$ and $0.1$. The idea is to start with a relaxed threshold in the early iterations where CMA-ES is exploring and gradually reduce it as iterations grow and CMA-ES starts exploring identified local minima. With this optimally tuned scheme, six points are added to the DoE. The found solution is $b^\ast = h^\ast = 239.12$ mm, has $0.28 \%$ discrepancy with the analytical solution. Note that the exact solution can be reached, should we increase the number of iterations of CMA-ES or refine the solution by a gradient-based algorithm. \figref{fig:Euler_appl} illustrates the convergence of CMA-ES. In total, only $18$ points were necessary to achieve convergence. By comparison, a one-shot approach with a DoE of size $18$ does not systematically converge to the reference solution. Additional points are needed most of the time.

\subsection{Two-dimensional problem}

This analytical example has been widely used for benchmark purposes in the related literature \citep{Du2004,Shan2008,Liang2004,DubourgThesis}. The optimization problem consists in minimizing the sum of the design parameters under three non-linear limit state functions whose deterministic formulation reads:
\begin{equation}\label{eq:1001}
\begin{split}
& \ve{d}^\ast = \arg \min_{\ve{d} \in \bra{0, 10}^2} d_1 + d_2  \quad \text{s.t.: }
\left\{ \begin{array}{ll}
\displaystyle{\mathfrak{g}_1 \prt{\ve{d}} = \frac{d_1^2 d_2}{20} - 1 \leq 0} \\[1em]
\displaystyle{\mathfrak{g}_2\prt{\ve{d}} = \frac{\prt{d_1 + d_2 - 5}^2}{ 30} + \frac{\prt{d_1 - d_2 - 12}^2}{120} - 1 \leq 0} \\[1em]
\displaystyle{\mathfrak{g}_3\prt{\ve{d}} = \frac{80} {\prt{d_1^2 + 8 d_2 + 5} - 1}} \leq 0
\end{array} \right..
\end{split}
\end{equation}
In order to solve the RBDO problem, we consider the following setting. The two design variables are considered as random: $X_i \sim \mathcal{N}\prt{d_i,0.6^2}, i = \acc{1,2}$. The target failure probability is $\bar{P}_{f_i} =1.35 \cdot 10^{-3}$ and thus corresponds to $\beta_i = 3$ for $i = \acc{1,2,3}$.

\begin{figure}[!ht]
	\begin{center}
          \subfloat[Convergence of the first stage of
          enrichment]{\includegraphics[width=0.49\textwidth]{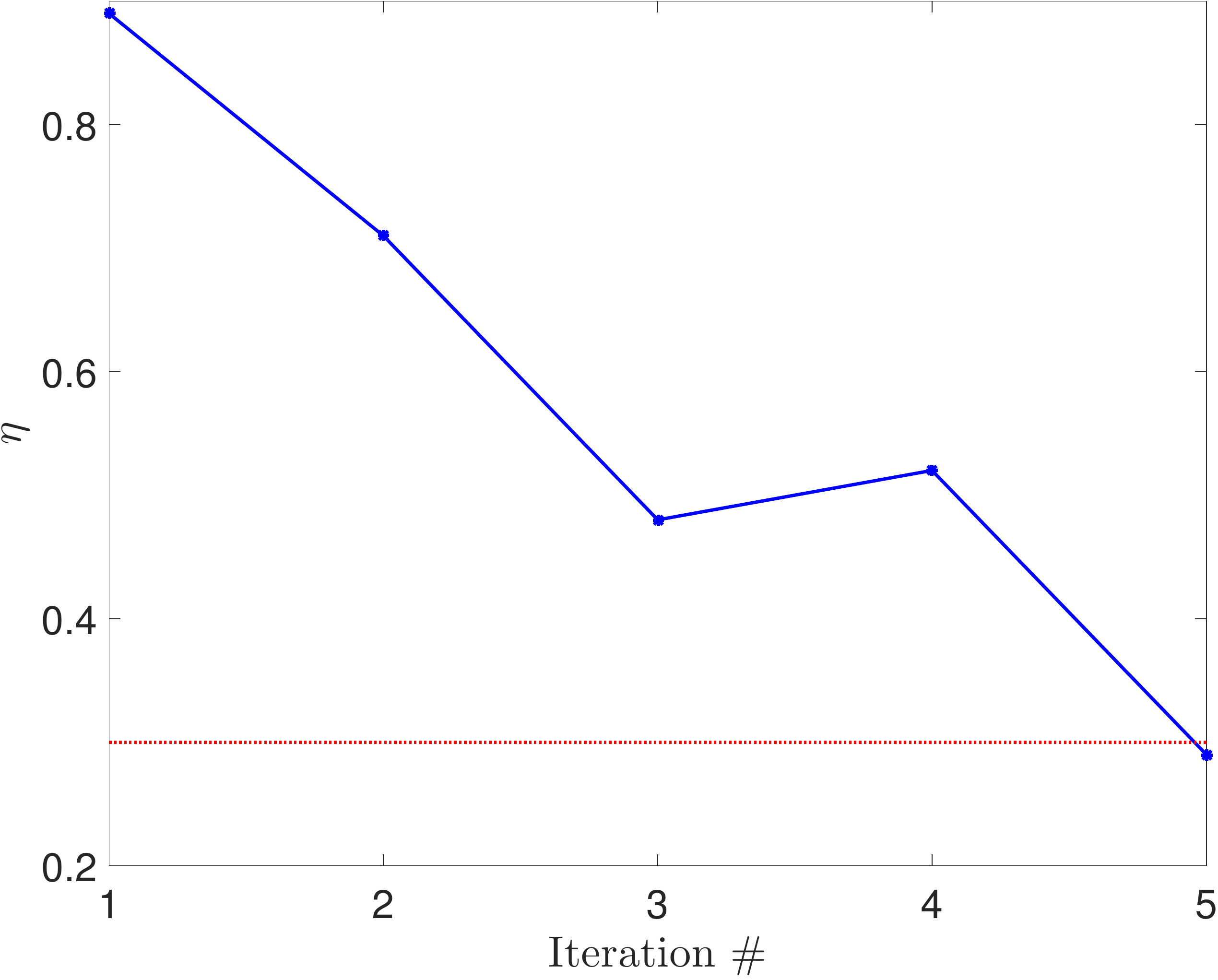}\label{fig:1001a}}%
          \hfill%
          \subfloat[Limit-state surface after the first stage of
          enrichment]{\includegraphics[width=0.49\textwidth]{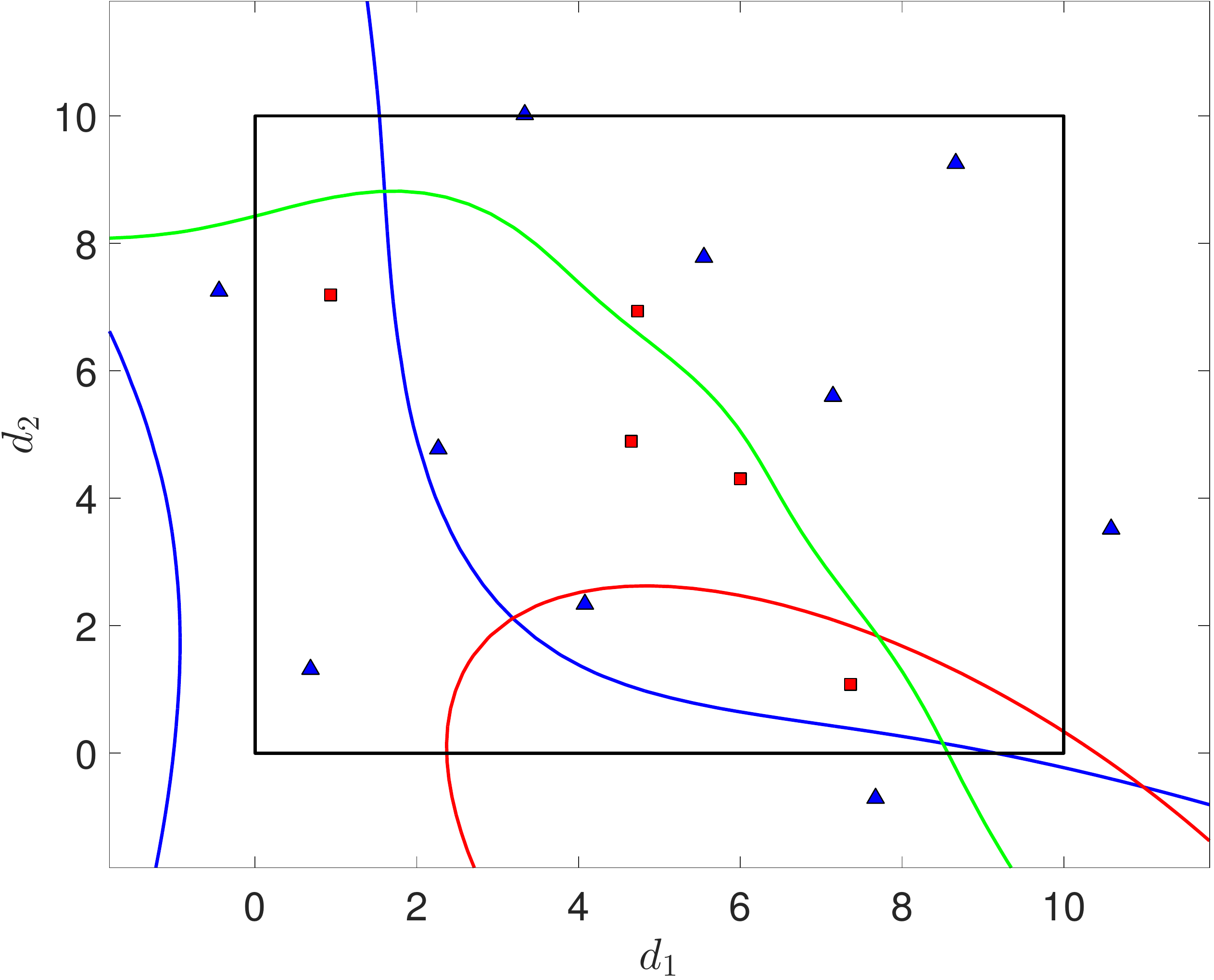}\label{fig:1001b}}%
		\caption{Illustration of the first stage of enrichment for the two-dimensional problem} %
		\label{fig:1001}%
	\end{center}
\end{figure}

We start the procedure with a $10$-point experimental design. Considering \\ $\bar{\eta}~=~0.3$, five points are added during the first stage of enrichment. \figref{fig:1001a} shows the convergence of this enrichment stage. In \figref{fig:1001b}, the contours of the limit-state with respect to the current Kriging models in the augmented space $\mathbb{X} = \bra{-1.8,11.8}^2$ are shown. In this figure, the black rectangle corresponds to the bounds of the design space $\mathbb{D} = \bra{0,10}^2$ and the initial and added points are shown respectively as blue triangles and red squares.

\begin{figure}[!ht]
	\begin{center}
          \subfloat[Evolution of the quantile accuracy
          criterion]{\includegraphics[width=0.5\textwidth]{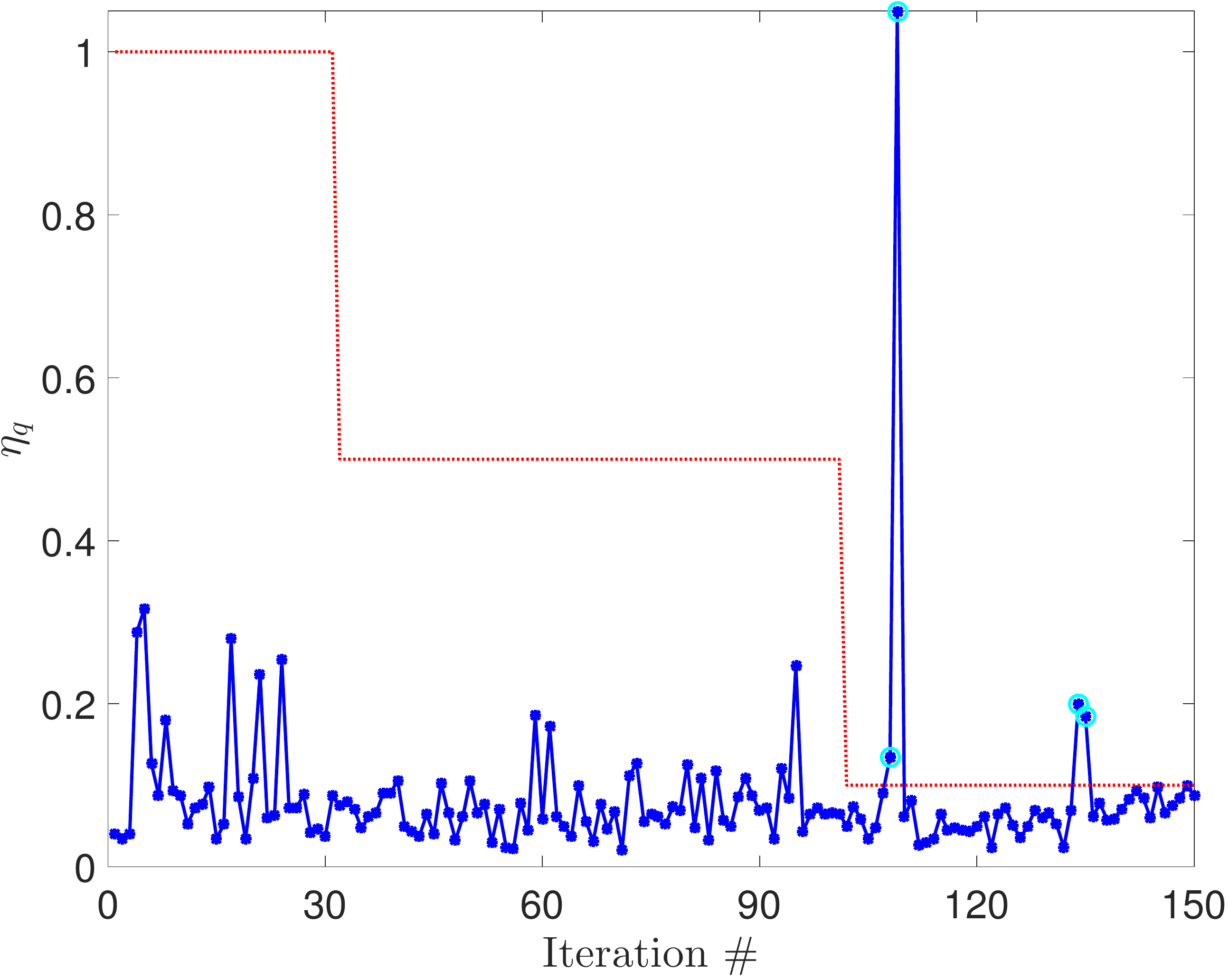}\label{fig:1002a}}%
          \hfill%
          \subfloat[Convergence of the constrained $(1+1)$-CMA-ES
          algorithm]{\includegraphics[width=0.49\textwidth]{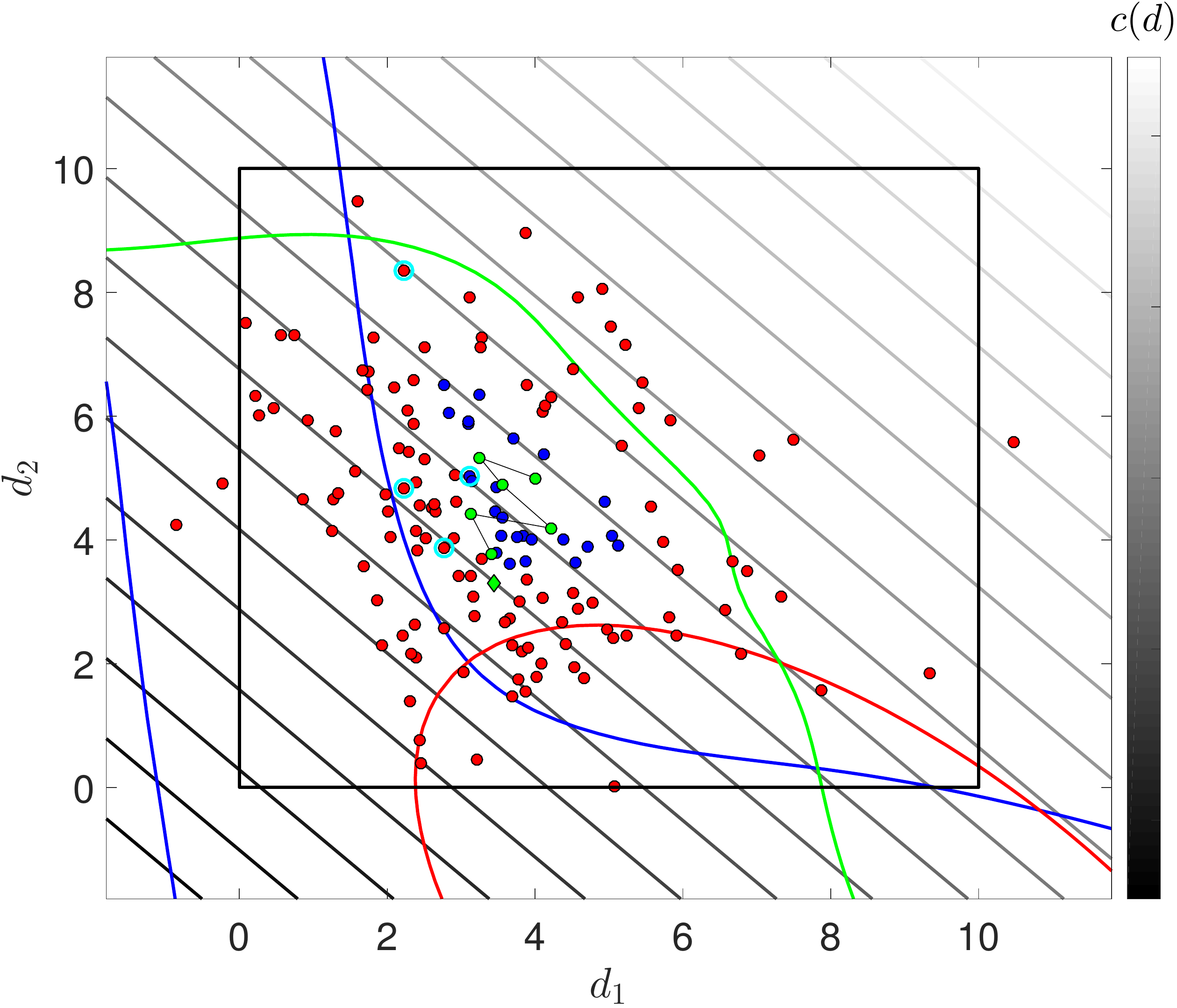}\label{fig:1002b}}%
		\caption{Illustration of the second stage of enrichment for the two-dimensional problem} %
		\label{fig:1002}%
	\end{center}
\end{figure}

We then proceed to optimization using constrained $(1+1)$-CMA-ES, starting from $\ve{d}^{(0)} = \acc{4, 5}^T$. The quantile accuracy thresholds are once more set in a simulated-annealing fashion as in the previous case with $\bar{\eta}_q = \acc{1,0.5,0.1}^T$. Convergence is achieved with four points added in the experimental design as illustrated in the diagnostic plots in \figref{fig:1002}. In the left panel, the evolution of the quantile accuracy criterion together with their associated thresholds are presented. The right panel illustrates convergence of CMA-ES algorithm. The red points violate the performance criteria (failure points). The blue and green ones are in the safe domain but only the latter improves the current best design during optimization.

Finally, \tabref{tab:1001} compares the results presented here with those reported in the literature for benchmark. All the selected methods provide a solution with a good accuracy. The difference rather lies in their costs. In these results, the single loop and the reliable design space (RDS) approaches require a relatively small number of functions evaluation despite they do not rely upon surrogate models. The two cases considering surrogate models (Meta-RBDO and Quantile-RBDO) are among the best in terms of model evaluations. In this example, the quantile-based approach we propose is on average the less expensive one. Since the initial design is random, the analysis is replicated $50$ times. The number of calls to the true models varies between $11$ and $23$, all of them leading to good results. On average the number of calls is $14.6$ and among them only three are above $20$.

\begin{table}[!ht]
	\centering
	\begin{threeparttable}
		\caption{Results comparison for the Choi problem.}
		\label{tab:1001}
		\begin{tabular}{lcccc}
			\hline
			{Method}    & {$d^\ast_1$}  & {$d^\ast_2$}  & {$\mathfrak{c}\prt{\ve{d}^\ast}$} & {$\mathfrak{g}$-calls} \\ \hline
			{Brute force}		      & {$3.45$}		& {$3.30$}	   & {$6.75$}	& {$\approx 10^6$}   \\
			{PMA$^1$}		      & {$3.43$}		& {$3.29$}	   & {$6.72$}	& {$1,551$}   \\
			{SORA$^2$}		      & {$3.44$}		& {$3.29$}	   & {$6.73$}	& {$151$}   \\
			{Single loop$^3$}		      & {$3.43$}		& {$3.29$}	   & {$6.72$}	& {$19$}   \\
			{RDS$^1$}		      & {$3.44$}		& {$3.28$}	   & {$6.72$}	& {$27$}   \\
			{Meta-RBDO$^4$}		      & {$3.46$}		& {$3.27$}	   & {$6.74$}	& {$20 (20/10/10)$}   \\
			{Quantile-RBDO}		      & {$3.44$}		& {$3.29$}	   & {$6.73$}	& {$17$}   \\  		 		 		
			\hline
		\end{tabular}
		\begin{tablenotes}
			\tiny
			\item $^1$ As calculated in \citet{Shan2008}
			\item $^2$ As calculated in \citet{Du2004}
			\item $^3$ As calculated in \citet{Liang2004}
			\item $^4$ As calculated in \citet{DubourgThesis}
		\end{tablenotes}
	\end{threeparttable}
\end{table}

\subsection{Bracket structure}

This mechanical example consists of the two-member bracket structure illustrated in \figref{fig:bra_str} \citep{Chateauneuf2008}. The two-members are pin-joined at the point B and a vertical load $P$ is applied on the right end of the member CD at a distance $L$ of its hinge. The aim is to minimize its weight while considering two failure modes:
\begin{figure}[!ht]
	\begin{center}
		\includegraphics[trim = 15mm 40mm 15mm 40mm, clip,width=0.75\textwidth]{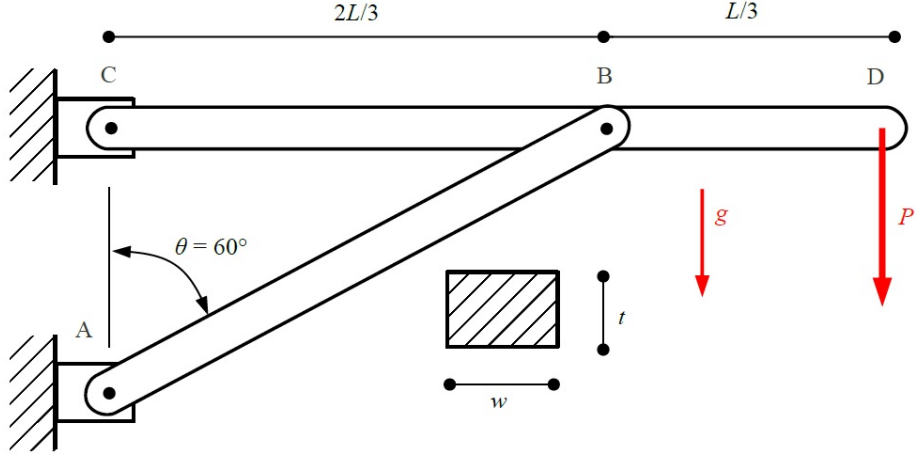}%
		\caption{A sketch of the bracket structure (as illustrated in \citet{DubourgThesis})} %
		\label{fig:bra_str}%
	\end{center}
\end{figure}
\begin{itemize}
	\item The bending stress in the member CD whose maximum value $\sigma_b$ is required to be smaller than the yield stress $\sigma_y$:
	\begin{equation}\label{eq:033}
	\mathfrak{g}_1 \prt{\ve{d},\ve{z}} = \sigma_y - \sigma_b,
	\end{equation}
	where  $\sigma_b = 6 M_B/w_{CD} t^2$, with $M_B = PL/3 + \rho g w_{CD} t L^2 / 18$. Here $w_{CD}$ and $t$ are the cross-sectional dimensions of CD, $\rho$ is the unit mass of its constitutive material and $g$ is the gravity intensity.
	\item The compression force $F_{AB}$ that must be lower than the critical Euler force $F_b$:
	\begin{equation}\label{eq:034}
	\begin{split}
	\mathfrak{g}_2 \prt{\ve{d},\ve{z}} = F_b - F_{AB} \quad \text{with:} \quad & F_b = \frac{\pi^2 E I}{L^{2}_{AB}} = \frac{\pi^2 E t w_{AB}^3}{12 \prt{2 L/3 sin \theta}^2}, \\
	& F_{AB} = \frac{1}{cos \theta} \prt{\frac{3P}{2} + \frac{3 \rho g w_{CD} t L}{4}}, \\	
	\end{split}
	\end{equation}
	where $w_{AB}$ and $L_{AB}$ are respectively the width and length of AB and $\theta$ is its inclination angle.
\end{itemize}

The RBDO application problem as in \citet{Chateauneuf2008} and \citet{Dubourg2011} consists in minimizing the weight of the structure, given the following objective function:
\begin{equation}\label{eq:035}
\mathfrak{c}\prt{\ve{d}} = \rho t L \prt{\frac{4 \sqrt{3}}{9} w_{AB} + w_{CD}},
\end{equation} 
where $\ve{d} = \acc{w_{AB},w_{CD},t}^T \in \mathbb{D} = \bra{5,30}^3$ represents the set of design parameters.

The target reliability index for this problem is set to $\beta_1 = \beta_2 = 2$, thus corresponding to a component failure probability of $0.0227$. The RBDO problem therefore reads:
\begin{equation}\label{eq:2001}
\begin{split}
& \ve{d}^\ast = \arg \min_{\ve{d} \in \mathbb{D}} \mathfrak{c}\prt{\vec{d}}  \quad \text{subject to: }
\left\{ \begin{array}{ll}
\displaystyle{\mathcal{P}\prt{\mathfrak{g}_1 \prt{\vec{X}\prt{\ve{d}},\ve{Z}} \leq 0} \leq \bar{P}_{f_1}} \\
\displaystyle{\mathcal{P}\prt{\mathfrak{g}_2 \prt{\vec{X}\prt{\ve{d}},\ve{Z}} \leq 0} \leq \bar{P}_{f_2},}
\end{array} \right.
\end{split}
\end{equation} 
where $\bar{P}_{f_i} = \Phi\prt{-\beta_1} \approx 0.0227$ and the functions $\mathfrak{g}_1$, $\mathfrak{g}_2$ and $\mathfrak{c}$ are respectively given by Eqs.~(\ref{eq:033}) -- (\ref{eq:035}).

The probabilistic model associated to this problem is shown in \tabref{tab:002}. \tabref{tab:002b} shows the bounds of the augmented space in which the training points are sampled. The surrogate model is built in the unit hypercube following a linear mapping from this augmented space.
\begin{table}[!ht]
	\centering
	\caption{Parameters of the variables defining the probabilistic model for the bracket structure problem: $\ve{d} = \acc{w_{AB}, w_{CD}, t}^T$ are the design variables and $\ve{z} = \acc{P,E,\sigma_y, \rho, L }^T$ are the environmental variables.}
	\label{tab:002}
	\begin{tabular}{lccc}
		\hline
		{Parameter}    & {Distribution}  & {Mean}  & {COV ($ \delta \%$)} \\ \hline
		{Width of AB ($w_{AB}$ in m)}		      & {Normal}		& {$w_{AB}$}	   & {$0.05$}	   \\
		{Width of CD ($w_{CD}$ in m)}	  & {Normal}	  & {$w_{CD}$}  & {$0.05$}   \\
		{Thickness ($t$ in m)}		  &{Normal}	  & {$t$}	 & {$0.05$}    \\
		{Applied load ($P$ in kN)}		  &{Gumbel}	  & {$100$}	 & {$0.15$}    \\
		{Young's modulus ($E$ in GPa)}		  &{Gumbel}	  & {$200$}	 & {$0.08$}    \\
		{Yield stress ($\sigma_y$ in MPa)}		  &{Lognormal}	  & {$225$}	 & {$0.08$}    \\
		{Unit mass ($\rho$ in kg/m$^3$)}		  &{Weibull}	  & {$7860$}	 & {$0.10$}    \\
		{Length ($L$ in m)}		  &{Normal}	  & {$5$}	 & {$0.05$}\\
		\hline
	\end{tabular}
\end{table}

\begin{table}[!ht]
	\centering
	\caption{Bounds of the augmented space for the bracket structure problem.}
	\label{tab:002b}
	\begin{tabular}{lccc}
		\hline
		{Parameter}    & {Lower bound}  & {Upper bound} \\ \hline
		{Width of AB ($w_{AB}$ in m)}		      & {$4.25$}		& {$34.5$}	\\
		{Width of CD ($w_{CD}$ in m)}	   & {$4.25$}		& {$34.5$}  \\
		{Thickness ($t$ in m)}		   & {$4.25$}		& {$34.5$}   \\
		{Applied load ($P$ in kN)}		  &{$15.98$}	  & {$109.58$}	   \\
		{Young's modulus ($E$ in GPa)}		  &{$110.38$}	  & {$224.31$}	\\
		{Yield stress ($\sigma_y$ in MPa)}		  &{$176.49$}	  & {$285.01$}	 \\
		{Unit mass ($\rho$ in kg/m$^3$)}		  &{$4760.20$}	  & {$9576.3$}	\\
		{Length ($L$ in m)}		  &{$4.25$}	  & {$5.75$}\\
		\hline
	\end{tabular}
\end{table}

\begin{figure}[!ht]
	\begin{center}
		\includegraphics[width=0.7\textwidth]{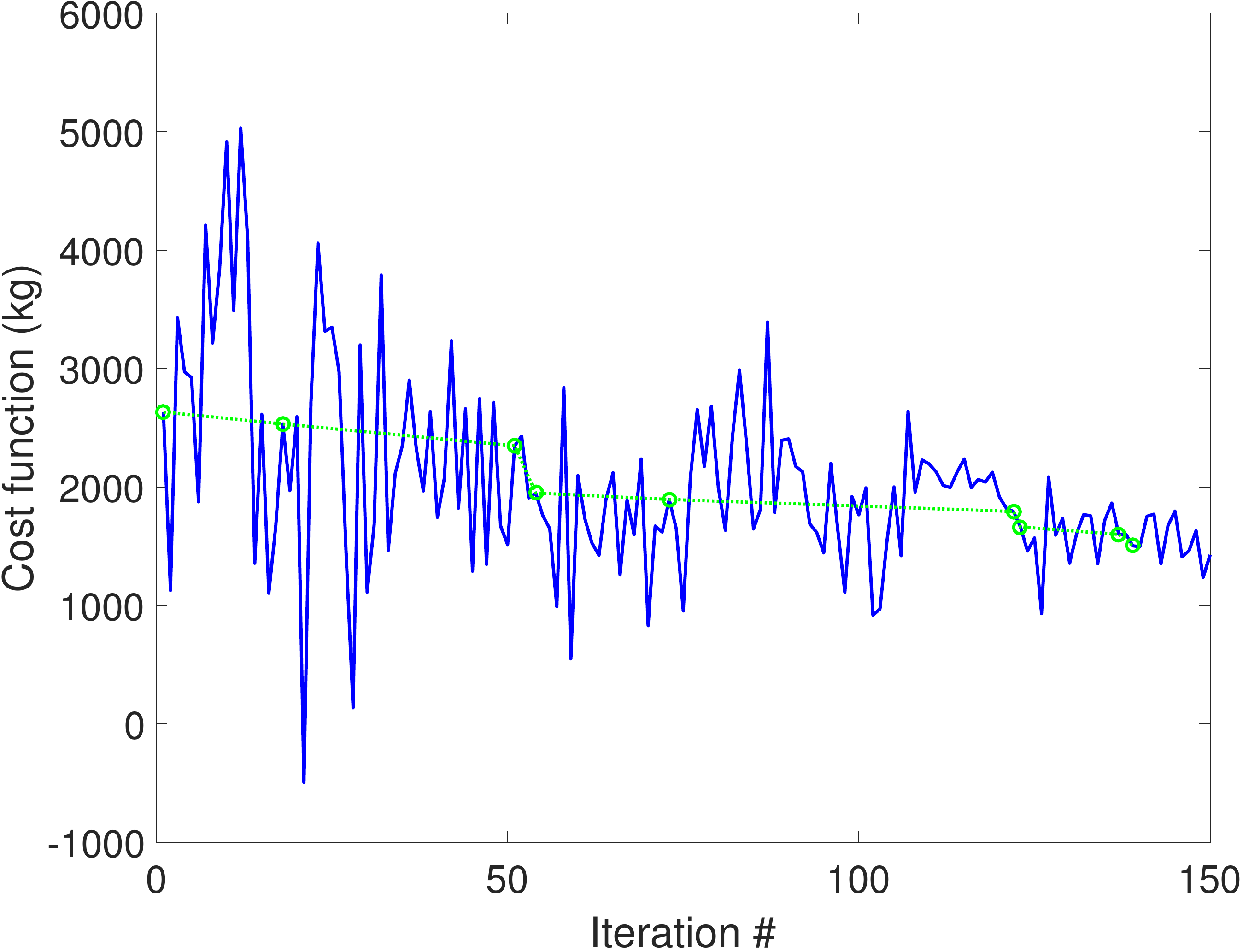}
		\caption{Convergence of the CMA-ES algorithm for the bracket structure.} %
		\label{fig:3001}%
	\end{center}
\end{figure}

For the solution of this problem, we start with an initial design of $50$ points. The threshold for the first stage of enrichment is set to $\bar{\eta} = 0.30$. A total of $60$ enrichments points have been added to reach the required accuracy through $6$~iterations of $10$ points each. The optimization is then initiated starting from $\ve{d}^{(0)} = \acc{6.1, 20.2, 26.9}^T$ which is also the initial design in the benchmark references \cite{Chateauneuf2008,DubourgThesis} and corresponds to the optimal deterministic solution. In the second stage, $K=3$ points are added per enrichment, thus leading to $15$ additional points in the DoE. Convergence of the CMA-ES algorithm is illustrated in \figref{fig:3001} where the evolution of the cost function with respect to the iteration number is shown. The green circles highlight the points that were feasible and improved the current best design. The CMA-ES algorithm is stopped after $150$ iterations and the solution locally refined through a gradient-based approach using the final Kriging model. The overall number of calls to the original model is $125$ for this illustrated case.

As in the previous example, we replicate the optimization $20$ times because of the random nature of the initial experimental design. The number of calls varies between $80$ and $170$, with the maximum clearly being an outlier. On average, this number of calls is $107$. In the light of this result, the proposed procedure is more efficient than the approaches from the two benchmark references as shown in \tabref{tab:003}. In this table, the brute force approach refers to a solution that is found by a quantile-based procedure directly relying on the true mechanical models rather than surrogates. Beside, the resulting weight saving is higher in our approach. This may be explained by the fact that we use a global optimization algorithm rather than a gradient-based one as was done in the two references.
\begin{table}[!ht]
	\centering
	\begin{threeparttable}
		\caption{Comparative results for the bracket structure. The PMA result comes from \cite{Chateauneuf2008} and Meta-RBDO from \cite{DubourgThesis}.}
		\label{tab:003}
		\begin{tabular}{lccccc}
			\hline
			{Design method}    & {Weight (kg)}  & {$w_{AB}$ (cm)}  & {$w_{CD}$ (cm)} & {$t$ (cm)} & {$\mathfrak{g}$-calls}\\ \hline
			{Brute force}		      & {$1357$}		& {$5.35$}	   & {$7.40$} & {$30.00$} & {$\approx 10^6$} \\		
			{PMA$^1$}		      & {$1673$}		& {$6.08$}	   & {$15.68$} & {$20.91$} & {$2340$}\\
			{Meta-RBDO$^2$}		      & {$1584$}		& {$5.80$}	   & {$12.80$} & {$23.30$} & {$160 (160/90)$}\\
			{Quantile-based RBDO}		      & {$1364$}		& {$5.57$}	   & {$7.28$} & {$30.00$} & {$125$} \\
			\hline
		\end{tabular}
		\begin{tablenotes}
			\scriptsize
			\item $^1$ As computed by \cite{Chateauneuf2008}
			\item $^2$ As calculated by \cite{DubourgThesis}
		\end{tablenotes}
	\end{threeparttable}
\end{table}

\subsection{Sidemember subsystem}

\begin{figure}[!ht]
	\begin{center}
          \subfloat[Parts of the sidemember
          subsystem]{\includegraphics[width=0.49\textwidth]{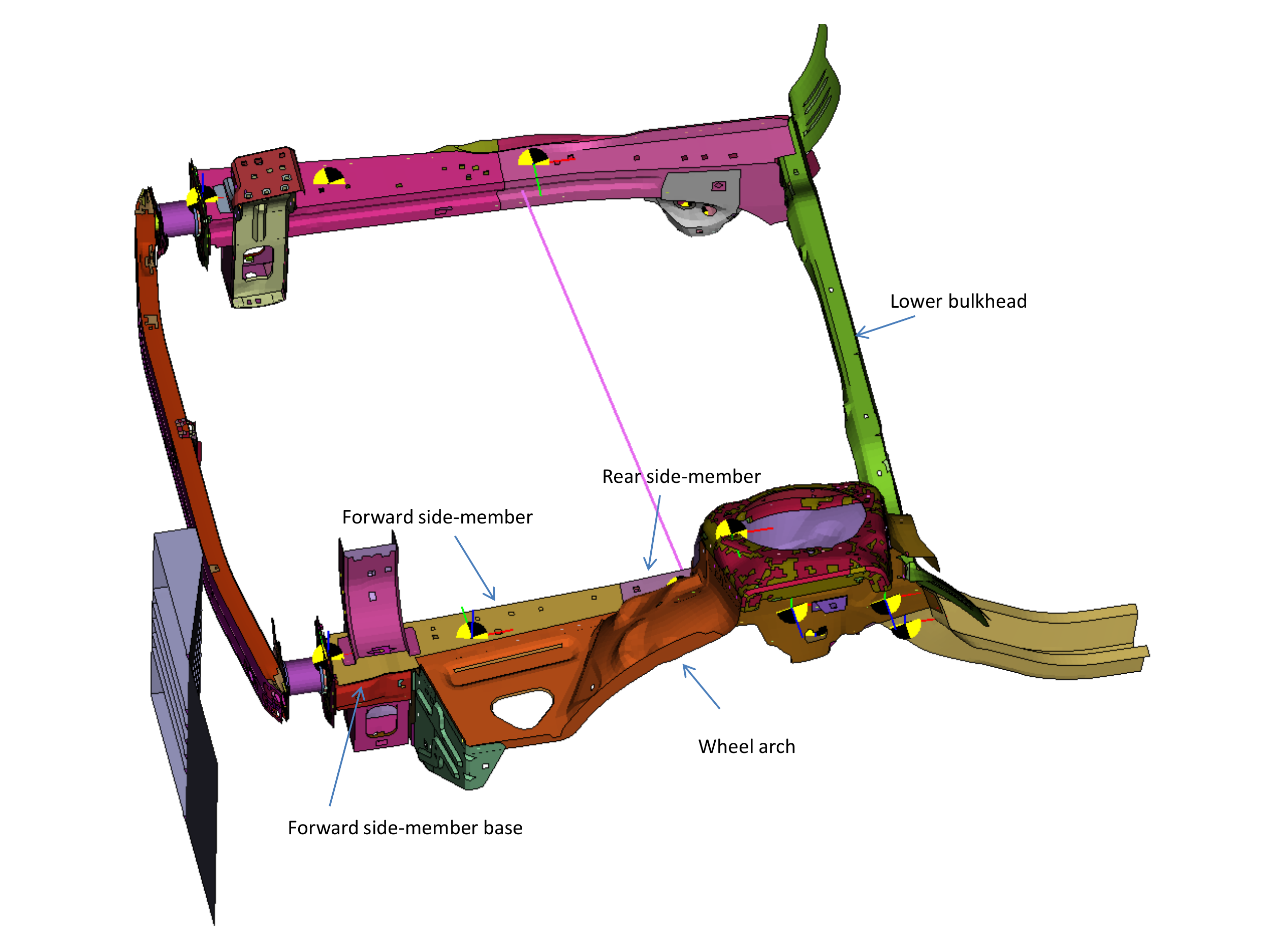}\label{fig:sid_sub_a}}%
          \hfill%
          \subfloat[A vehicle under frontal
          impact]{\includegraphics[width=0.49\textwidth]{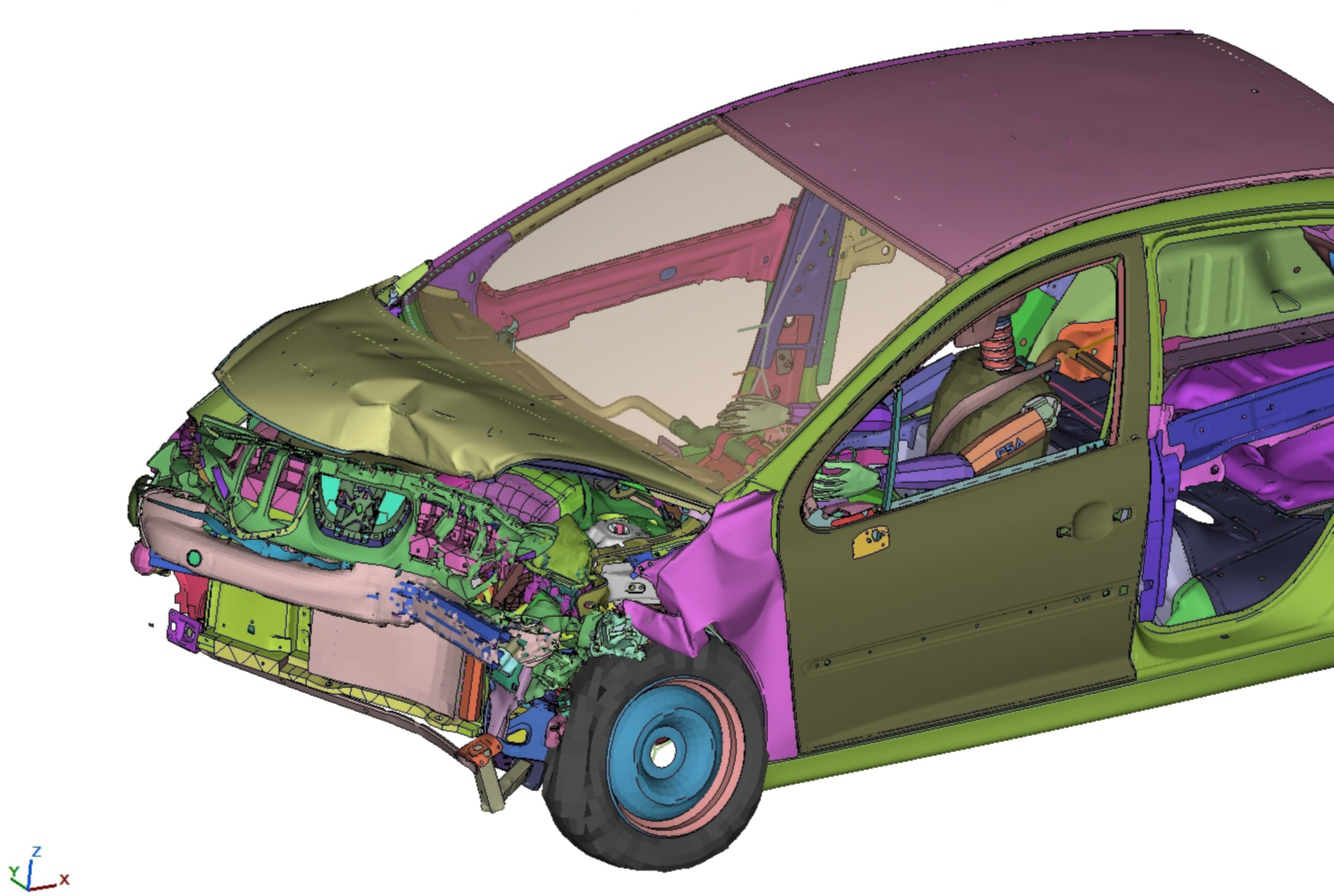}\label{fig:Sid_sub_b}}%
		\caption{Sidemember subsystem} %
		\label{fig:Sid_sub}%
	\end{center}
\end{figure}

This final application is related to the lightweight design of an automotive body structure under crashworthiness constraints. This involves finding the best distribution of the metal sheet thicknesses which allows one to satisfy frontal impact-related constraints. These constraints are evaluated by finite element crash simulations which are extremely time-consuming, \ie $24$ hours for a single model run on distributed CPUs. The use of surrogate models is therefore the only alternative in order to perform such an optimization. In this application, we consider the so-called \emph{sidemember subsystem} which is a subset of the front end of a vehicle. This subsystem actually has the same behavior in frontal impact as a full vehicle, yet requires reduced computational time ($10$ to $15$ minutes on a cluster of $48$ CPUs). The sidemember subsystem is illustrated in \figref{fig:Sid_sub}. Five parts are considered for optimization as shown in the figure. To account for noise which is inherent to frontal impact, some parameters of the crash protocol are considered as random. These are respectively the initial speed and the position of the barrier: $V \sim \mathcal{U}\prt{34,35}$ km/h and $P \sim \mathcal{N}\prt{0,2}$ mm. The two constraints that are considered for this problem are the maximum wall contact force that should not be larger than $\bar{\mathfrak{g}}_1 = 170$ kN and the maximum sidemember compression which should be kept below $\bar{\mathfrak{g}}_2 = 525$ mm. In order to obtain a conservative design with respect to uncertainties, the quantile-based optimization procedure is applied with a quantile level $\alpha_1 = \alpha_2 = 0.95$. The associated RBDO problem reads as follows:
\begin{equation}\label{eq:2002}
\begin{split}
& \ve{d}^\ast = \arg \min_{\ve{d} \in \mathbb{D}} \mathfrak{c}\prt{\vec{d}}  \quad \text{subject to: }
\left\{ \begin{array}{ll}
\displaystyle{\mathcal{P}\prt{\bar{\mathfrak{g}}_1 - \mathcal{M}_1 \prt{\ve{d},\ve{Z}} \leq 0} \leq 0.05 }\\
\displaystyle{\mathcal{P}\prt{\bar{\mathfrak{g}}_2 - \mathcal{M}_2 \prt{\ve{d},\ve{Z}} \leq 0} \leq 0.05,}
\end{array} \right.
\end{split}
\end{equation} 
where $\ve{Z} = \acc{V,P}^T$ and  $\mathcal{M}_1$ and $\mathcal{M}_2$ are the outputs of the finite element model giving respectively the maximum wall force and maximum sidemember compression.

The initial design and the bounds of the augmented space associated to this problem are given in \tabref{tab:004}. Since the design variables are deterministic $\mathbb{X}$ reduces to the design space.
\begin{table}[!ht]
	\centering
	\caption{Bounds of the augmented space and initial design for the sidemember subsystem.}
	\label{tab:004}
	\begin{tabular}{lccccccc}
		\hline
		{Param.} & {$d_1$ (mm)} & {$d_2$ (mm)} & {$d_3$ (mm)} & {$d_4$ (mm)} & {$d_5$ (mm)} & {$V$ (km/h)} & {$P$ (mm)} \\ \hline
		{Lower} & {$1.5$} & {$1.5$} & {$2$} & {$1.5$} & {$0.6$} & {$34$} & {$-6$} \\
		{Upper} & {$2.5$} & {$2.5$} & {$3$} & {$2.5$} & {$1.2$} & {$35$} & {$6$} \\
		{Initial} & {$2$} & {$2$} & {$2.5$} & {$2$} & {$0.9$} & {$-$} & {$-$} \\
		\hline
	\end{tabular}
\end{table}

\begin{figure}[!ht]
	\begin{center}
          \includegraphics[width=0.6\textwidth]{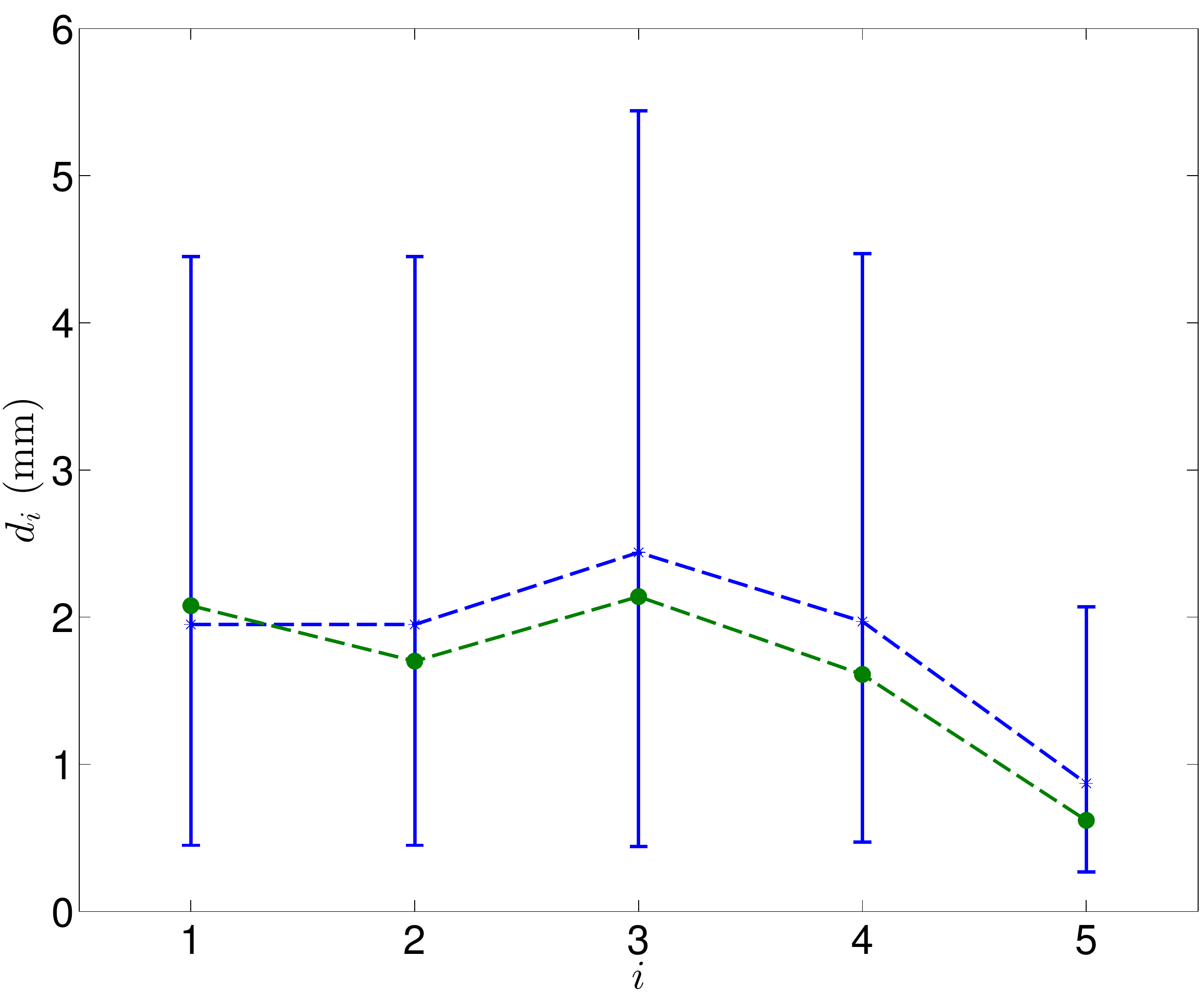}
		\caption{Comparison of the initial and optimal designs with respect to each parameter for the $5$-part sidemember subsystem. The blue color stands for the nominal design and the green color for the optimal one.}%
		\label{fig:Sous_syst_df}%
	\end{center}
\end{figure}

To solve this seven-dimensional highly non-linear problem, we start with
an initial design of $70$ points. In the first stage of enrichment, $20$
points are added during two iterations thus leading to a global accuracy
criterion $\eta \leq 0.2$. For the second stage, we set $\eta_q = 0.01$,
thus accepting a $1 \%$ relative error. To keep the enrichment to what
is strictly necessary, we decide to enrich only around designs that
improve the current best ones. In this way, eight iterations with $K =
3$ simultaneously added points were carried out. The overall number of
calls to the finite element model therefore amounts to only $114$.
At convergence, the found solution results to a weight saving of $1.08$
kg, that is $11.5 \%$ of the initial weight, which is considered
significant in car manufacturing. The thicknesses associated to the
initial and optimal solutions are shown in \figref{fig:Sous_syst_df}.
The corresponding quantile constraints for this solution are
$\widehat{\mathfrak{q}}_{\alpha_1}\prt{\ve{d}^\ast} = 155.62$ kN and
$\widehat{\mathfrak{q}}_{\alpha_2}\prt{\ve{d}^\ast} = 523.12$ mm, which are below
the thresholds.

The validation of the reliability of this solution with respect to the finite element model is not possible due to the large cost of a single run. Instead, we focus here on the accuracy of the Kriging surrogates in the vicinity of the optimal design. To this end, we estimate quantiles with the original and surrogate models from a set of Monte Carlo samples of size $100$.  To account for this reduced size, we consider the mean value of the quantiles estimates obtained from $500$ bootstrap replicates. Each bootstrap replicate consists in sampling with replacement $100$ points from the available data. The resulting relative error is still biased because of the small size of the Monte Carlo set. However, this allows us to give a flavor on the ability of the surrogate model to approximate the quantile in the vicinity of the found solution. \tabref{tab:005} compares the results by considering the finite element model ($\widehat{\mathfrak{q}}_{\alpha}^{\text{FE}}$) on the one hand and the Kriging model ($\widehat{\mathfrak{q}}_{\alpha}^{\text{KRG}}$) on the other. The two responses are quite close for each output, showing that the Kriging models were accurate enough (at least locally) for the purpose of quantile estimation.
\begin{table}[!ht]
	\centering
	\caption{Quantiles of the performance criteria $\mathfrak{g}_1$
		and $\mathfrak{g}_2$ computed from the Kriging model
		($\widehat{\mathfrak{q}}_{\alpha}^{\text{KRG}}$) (resp. the original model ($\widehat{\mathfrak{q}}_{\alpha}^{\text{FE}}$)) obtained from $100$
		Monte Carlo samples, averaged over $500$ bootstrap replicates.}
	\label{tab:005}
	\begin{tabular}{lcccc}
          \hline
          {Criterion} & {$\mathfrak{g}_1$ (kN)} & {$\mathfrak{g}_2$ (mm)}\\ \hline
          {Original model $\widehat{\mathfrak{q}}_{\alpha}^{\text{FE}}$} & {$150.66$} & {$527.81$} \\
          {Kriging model $\widehat{\mathfrak{q}}_{\alpha}^{\text{KRG}}$}		& {$148.02$}	 & {$528.04$}  \\
          {Error ($\%$)} & {$1.75$} & {$0.04$} \\
          \hline
	\end{tabular}
	
\end{table}

\section{Conclusion}
The aim of this paper is to propose a quantile-based, conservative
optimization procedure for structures in an uncertain environment.
Furthermore, structures whose behavior is simulated by high-fidelity and
expensive-to-evaluate models are considered. Such simulations are
time-consuming. Surrogate modeling approaches are therefore introduced
as computationally costless approximations of these models.

The optimization problem is first posed in the framework of
reliability-based design optimization (RBDO). After a brief review of
the most-widely used techniques to solve a RBDO problem, we formulate a
new quantile-based approach of optimal design. This approach is
motivated by the relatively high target failure probabilities that can
be accepted in the applications under consideration in the field of car
body design.  These probabilities of failure will be estimated by crude
Monte Carlo sampling.

Kriging, with its basic equations, is introduced for the purpose of
surrogate modeling. To further reduce the computational burden
associated to building the Kriging surrogate model, a two-stage
enrichment of the design of computer experiments is proposed in an
augmented space that combines both design variables and uncertain
environmental variables. The first stage, which is global, aims at
reducing the overall Kriging epistemic uncertainty by adding points in
the vicinity of the limit-state surfaces. The second stage, which is
local, is embedded in the optimization procedure. At each iteration, the
accuracy of the estimated quantiles is checked. Enrichment of the design
of experiments is made locally only when the accuracy is not sufficient.
This allows us to direct the experimental design points to regions of
the space that decrease significantly the cost function while ensuring
that the performance criteria are fulfilled.

Three applications are considered to validate the proposed procedure. The first one is a five-dimensional example related to a beam buckling problem, whose analytical solution can be computed. This allows us to validate the proposed method against exact results. The second and third problems respectively involve three non-linear limit state functions and a bracket structure. The optimal solutions obtained from different approaches are already available in the literature. The application of the proposed procedure shows increased efficiency as the number of calls to the original computational model is reduced. For the bracket structure, a better solution in terms of the cost function is even found compared to the best results available in the literature. Finally, we apply the methodology to an industrial problem related to the lightweight design of an automotive sidemember subsystem under frontal impact. A reliable solution is found within a reasonable number of calls to the expensive finite element model. All these applications feature relatively low-dimensional problems. Applications to high-dimensional cases, say $s > 20$, is still a challenging task and require further work. 

%%%%%%%%%%%%%%%%%%%%%%%%%%%%%%%%%%%%%%%%%%%%%%%%%%%%%%%%%%%%%%%%%%%%%%%%%%%%
%% References
\nocite{SudretHDR}
\bibliographystyle{chicago}
\bibliography{biblioMM}
\end{document}